\documentclass[aps,prd,onecolumn,groupedaddress,showpacs,nofootinbib,amssymb]{revtex4}
\usepackage[dvips]{graphicx}
\usepackage{amssymb}
\usepackage{amsmath}
\usepackage{graphicx,,color}
\usepackage{amsfonts}
\usepackage{bm}
\usepackage{cancel}
\usepackage{comment}
\usepackage{tablefootnote}

\newcommand\be{\begin{equation}}
\newcommand\ee{\end{equation}}

\allowdisplaybreaks[4]

\begin{document}

\tolerance=5000

\title{$f(R)$ Gravity Phase Space in the Presence of Thermal Effects}
\author{V.K.~Oikonomou,$^{1,2}$\,\thanks{v.k.oikonomou1979@gmail.com}
F.P.~Fronimos,$^{1}$\,\thanks{fotisfronimos@gmail.com}
N.Th.~Chatzarakis,$^{1}$\,\thanks{nchatzar@physics.auth.gr}}
\affiliation{$^{1)}$ Department of Physics, Aristotle University
of Thessaloniki, Thessaloniki 54124,
Greece\\
$^{2)}$ Laboratory for Theoretical Cosmology, Tomsk State
University of Control Systems and Radioelectronics, 634050 Tomsk,
Russia (TUSUR)\\
}

\tolerance=5000

\begin{abstract}
In this paper, we shall consider $f(R)$ gravity and its
cosmological implications, when an extra matter term generated by
thermal effects is added by hand in the Lagrangian. We formulate
the equations of motion of the theory as a dynamical system, that
can be treated as an autonomous one only for specific solutions
for the Hubble rate, which are of cosmological interest.
Particularly, we focus our analysis on subspaces of the total
phase space, corresponding to (quasi-)de Sitter accelerating
expansion, matter-dominated and radiation-dominated solutions. In
all the aforementioned cases, the dynamical system is an
autonomous dynamical system. With regard to the thermal term
effects, these are expected to significantly affect the evolution
near a Big Rip singularity, and we also consider this case in
terms of the corresponding dynamical system, in which case the
system is non-autonomous, and we attempt to extract analytical and
numerical solutions that can assess the specific cases. This
course is taken twice: the first for the vacuum theory and the
second when two perfect fluids (dust and radiation) are included
as matter sources in the field equations. In both cases, we reach
similar conclusions. The results of this theory do not differ
significantly from the results of the pure $f(R)$ in the de Sitter
and quasi-de Sitter phases, as the same fixed points are attained,
so for sure the late-time era de Sitter is not affected. However,
in the matter-dominated and radiation-dominated phases, the fixed
points attained are affected by the presence of the thermal term,
so surely the thermal effects would destroy the matter and
radiation domination eras. However, with regard to the Big Rip
case, several instabilities are found in the dynamical system,
since the initial conditions dramatically affect the behavior of
the dynamical system near the starting point of the $e$-foldings
number evolution. Finally, we consider the effects of the thermal
term on the late-time evolution of the Universe, by examining
several statefinder quantities. Our findings indicate that these
are reportable only under extreme fine tuning of the
multiplicative coefficient of the thermal term.
\end{abstract}


\maketitle

\section{Introduction}

Undoubtedly dark energy is one of the current mysteries in
theoretical physics. The observation of the late-time acceleration
driven by the mysterious dark energy in the late 90's
\cite{Riess:1998cb}, still remains a firm proof that the Universe
deviates in a major degree from the standard cosmological model
described by general relativity. Modified gravity seems to harbor
in a quite successful way the consistent description of the dark
energy era
\cite{reviews1,reviews2,reviews3,reviews4,reviews5,reviews6} and
in some cases it is also possible to describe under the same
theoretical framework the inflationary era and the dark energy era
\cite{Odintsov:2020nwm,Nojiri:2003ft}. Nevertheless, until the
generator of this dark energy fluid is found, current research
also focuses on side effects related to the dark energy era. One
first study in this respect, focused on the thermal effects caused
in the Friedmann equation, of the form $\alpha H^4$ studied in
Ref. \cite{Nojiri:2020sti}. In this work we shall use the
dynamical system approach
\cite{Bahamonde:2017ize,Oikonomou:2019muq,Oikonomou:2019nmm,Odintsov:2018uaw,Odintsov:2018awm,Boehmer:2014vea,Bohmer:2010re,Goheer:2007wu,Leon:2014yua,Guo:2013swa,Leon:2010pu,deSouza:2007zpn,Giacomini:2017yuk,Kofinas:2014aka,Leon:2012mt,Gonzalez:2006cj,Alho:2016gzi,Biswas:2015cva,Muller:2014qja,Mirza:2014nfa,Rippl:1995bg,Ivanov:2011vy,Khurshudyan:2016qox,Boko:2016mwr,Odintsov:2017icc,Granda:2017dlx,Landim:2016gpz,Landim:2015uda,Landim:2016dxh,Bari:2018edl,Ganiou:2018dta,Shah:2018qkh,Oikonomou:2017ppp,Odintsov:2017tbc,Dutta:2017fjw,Odintsov:2015wwp,Kleidis:2018cdx,Oikonomou:2019boy,Chatzarakis:2019fbn}.
in order to reveal the effects of these thermal terms in subspaces
of the total phase space of the cosmological system, with special
cosmological interest. Our study will focus on $f(R)$ theories of
gravity in vacuum or the presence of perfect matter fluids, and
our aim is to reveal the structure of the phase space subspaces
corresponding to (quasi) de Sitter, matter dominated or radiation
dominated eras in the presence of an added by hand thermal term.
We shall formulate the equations of motion of the theory as a
dynamical system, that can be rendered as an autonomous dynamical
system only for specific solutions for the Hubble rate, which are
of cosmological interest. Specifically, we shall be interested in
subspaces of the total phase space, corresponding to (quasi-)de
Sitter accelerating expansion, matter-dominated and
radiation-dominated solutions as we already mentioned, in which
cases, the dynamical system can be rendered an autonomous
dynamical system. With regard to the thermal term effects, these
are expected to affect the evolution near a Big Rip singularity,
and we also consider this case in terms of the corresponding
dynamical system, in which case the system is non-autonomous, and
we attempt to extract analytical and numerical solutions that can
assess the specific cases. We perform all the aforementioned
studies for both the vacuum theory and in the presence of perfect
matter fluids (dust and radiation). In both cases, we reach
similar conclusions. The results of this theory do not differ
significantly from the results of the pure $f(R)$ in the de Sitter
and quasi-de Sitter phases, as the same fixed points are attained.
However, in the matter-dominated and radiation-dominated phases,
the fixed points which we found are affected significantly by the
presence of the thermal term, thus the matter and radiation
domination eras are destroyed essentially by thermal effects. This
clearly shows that these thermal terms should not be in the
Lagrangian during these eras, as was expected. What we expected
though is that the dynamical system would be strongly unstable for
the Big Rip singularity, and our analysis exactly verified this
result. Actually, in the Big Rip case, the dynamical system is
strongly unstable, and the initial conditions seem to dramatically
affect the behavior of the dynamical system near the starting
point of the $e$-foldings number evolution. Finally, we consider
the effects of the thermal term on the late-time evolution of the
Universe (present time acceleration), by concretely examining
several statefinder quantities, using a numerical approach. Our
findings indicate that these are significant only under extreme
fine tuning of the multiplicative coefficient $\alpha$ of the
thermal term $\alpha H^4$. Thus with our results we verified that
the thermal effects should only affect the cosmological evolution
of $f(R)$ gravity near a Big Rip singularity and should not be
present during the matter and radiation domination eras, since
they would destroy all the desirable evolution properties during
these deceleration eras. Also during the acceleration eras,
materialized by (quasi) de Sitter fixed points, the thermal
effects are insignificant, as the dynamical system study shows, or
can affect the large redshift ($z\sim 6$) dark energy
singularities, under extreme fine tuning, at least when viable
models of $f(R)$ gravity are used.

The paper is organized as follows. In section II, we examine the
case of the vacuum $f(R)$ gravity with an additional fluid term
for the classical cosmological eras, whereas it can be modelled
and analyzed as an autonomous dynamical system. The non-autonomous
case is assessed in section III, where the finite-time
singularities are considered. These two cases are considered again
in the same manner in section IV where two perfect fluids are also
included as matter sources in the field equations of the $f(R)$.
In section V we study quantitatively the effects of the thermal
terms on the late-time era by examining several late-time
statefinder quantities. Finally, the conclusions follow at the end
of the paper.

\section{The Case of the Vacuum $f(R)$ Autonomous System}

It is known that near a Big Rip singularity, thermal effects may
affect the cosmological evolution \cite{Nojiri:2020sti}, since
these might become significant due to the cosmological horizon
``surrounding'' the singularity. In this research line, we shall
study in some detail the effects of the thermal terms on the
dynamical system of $f(R)$ gravity. Let us commence our study by
introducing the gravitational action and we shall consider the
vacuum $f(R)$ gravity in the presence of thermal correction terms
in the equation of motion, in which case the action is,
\begin{equation}
\centering \label{action} S=\int{\mathrm{d^4 x}\sqrt{-g}\left(
\frac{f(R)}{2\kappa^2} +\mathcal{L}_{eff} \right)}\, ,
\end{equation}
where $g$ is the determinant of the metric tensor $g^{\mu\nu}$,
$\kappa=\dfrac{1}{M_P}$ with $M_P$ being the reduced Planck mass,
$f(R)$ is an arbitrary for the time being function of the Ricci
scalar, which for a flat Friedmann-Robertson-Walker (FRW)
background is given by the expression $R=6(2H^2+\dot H)$ due to
the flat cosmological background assumed. It is written as a
function depending solely on Hubble's parameter $H=\dfrac{\dot
a}{a}$ where as usual $a(t)$ denotes the scale factor and the
``dot'' implies differentiation with respect to cosmic time $t$.
Lastly $\mathcal{L}_{eff}$ is an additional term which quantifies
the thermal effects on the late-time era. Its influence in the
equations of motion will be shown explicitly in the equations of
motion that now follow. In particular, by implementing the
variation principle with respect to the metric tensor, one obtains
the following equations of motion,
\begin{equation}
\centering
\label{motion1}
0=-\frac{f(R)}{2}+3(\dot H+H^2)F(R)-18(4\dot HH^2+H\ddot H)\frac{dF}{dR}+\alpha\kappa^2H^4\, ,
\end{equation}
where $F=\dfrac{\mathrm{d}f}{\mathrm{d}R}$. The extra term
$\alpha\kappa^2H^4$ is produced by the effective fluid Lagrangian
density $\mathcal{L}_{eff}$ that quantifies the thermal effects in
Eq. (\ref{action}) and its influence on the dynamics of the total
cosmological fluid is the main subject of this paper. In order to
form an appropriate dynamical system from the equations of motion,
and study in detail the dynamics of the system, we introduce the
following dimensionless dynamical variables,
\begin{align}
\centering
\label{xvariables}
x_1&=-\frac{\dot F}{HF}&x_2&=-\frac{f}{6FH^2}&x_3&=\frac{R}{6H^2}&x_4&=\frac{\alpha\kappa^2H^2}{3F}\, ,
\end{align}
The first three variables are identical with the ones used in the
case of vacuum $f(R)$ gravity \cite{Odintsov:2017tbc}, hence the
only new variable is the dynamical variable $x_4$. Genuinely
speaking, it is introduced and chosen in this way on purpose, in
order to rewrite Eq. (\ref{motion1}) in terms of the dynamical
variables elegantly as follows,
\begin{equation}
\centering \label{motion3} x_1 + x_2 + x_3 + x_4 = 1 \, .
\end{equation}
Let us now proceed with the construction of the dynamical system,
which is non-linear in our case. In the following, we shall make
use of the $e$-foldings number as a dynamical variable, instead of
the cosmic time, in order to designate a new differential operator
so as to replace the time derivatives. In principle, since
$H=\dfrac{\mathrm{d}N}{\mathrm{dt}}$, the differential operator
reads,
\begin{equation}
\centering
\label{dN}
\frac{\mathrm{d}}{\mathrm{dN}}=\frac{1}{H}\frac{\mathrm{d}}{\mathrm{dt}}\, ,
\end{equation}
In consequence, the non-linear dynamical system takes the
following form,
\begin{equation}
\centering
\label{dx1}
\frac{\mathrm{d}x_1}{\mathrm{dN}}= x_1(x_1+3-x_3) +2(x_3-2)(1-2x_4) \, ,
\end{equation}
\begin{equation}
\centering
\label{dx2}
\frac{\mathrm{d}x_2}{\mathrm{dN}}= x_2(x_1-2x_3+4) -4(x_3-2) +m \, ,
\end{equation}
\begin{equation}
\centering
\label{dx3}
\frac{\mathrm{d}x_3}{\mathrm{dN}}= -m -2(x_3-2)^2 \, ,
\end{equation}
\begin{equation}
\centering
\label{dx4}
\frac{\mathrm{d}x_4}{\mathrm{dN}}= x_4(x_1+2x_3-4) \, .
\end{equation}
As expected, for $x_4\to 0$, one obtains the usual dynamical
system for a pure $f(R)$ gravity in vacuum. In addition, we
introduced a new parameter $m=-\dfrac{\ddot H}{H^3}$ which is of
crucial importance in the following sections, as we now explain.
Apparently, for a general solution for the Hubble rate $H(t)$, the
parameter $m$ is time-dependent and hence, the dynamical system
(\ref{dx1})-(\ref{dx4}) is rendered non-autonomous. However, for
several cosmological solutions of interest, the parameter $m$
takes constant values. Specifically, if the Hubble rate describes
a quasi-de Sitter or a de Sitter evolution, or a matter or even a
radiation dominated evolution, the parameter $m$ takes constant
values. Specifically, these values will not be random but rather
we shall showcase that three choices, namely $m=0$ and
$m=-\dfrac{9}{2}$ and $m=-8$ result in decent descriptions of the
inflationary era, the matter-dominated era and the
radiation-dominated era respectively. These values are the same
irrespective of the presence of the thermal effects generated
terms. By assigning specific values on the parameter $m$ we
actually investigate a subspace of the total phase space
corresponding to the dynamical system. Practically, we do not fix
the Hubble rate, but we rather investigate the behavior of the
dynamical system for specific values of the parameter $m$, which
in turn indeed correspond to several solutions of cosmological
interest, thus we divide by hand the total phase space of the
dynamical system, which by the way is impossible to study in a
concrete way, to subspaces of cosmological interest. Our aim in
this paper is to study what are the effects of the thermal terms
$\sim H^4$, or equivalently of the variable $x_4$, on the
dynamical system, focusing on the subspaces of the total phase
space corresponding to the values of the parameter $m$ describing
the quasi-de Sitter or de Sitter ($m=0$), the matter dominated era
($m=-\dfrac{9}{2}$) and the radiation dominated era ($m=-8$).

For a general constant value of the parameter $m$, there exist
equilibria which are somewhat symmetric. In particular, the
aforementioned fixed points are,
\begin{align}
\centering \label{equilibrium} &\phi_*^1 =\left( \frac{1}{4}
\left(-\sqrt{-2m} -2 -\sqrt{4 -2m +20\sqrt{-2m}}\right),
\frac{1}{4} \left(3\sqrt{-2m} -2 +\sqrt{4 -2m +20
\sqrt{-2m}}\right), 2-\frac{\sqrt{-2m}}{4}, 0 \right) \\ \notag
&\phi_*^2 =\left( \frac{1}{4} \left(-\sqrt{-2m} -2 +\sqrt{4 -2m
+20\sqrt{-2m}}\right), \frac{1}{4} \left(3 \sqrt{-2m} -2 -\sqrt{4
-2m +20 \sqrt{-2m}}\right), 2-\frac{\sqrt{-2m}}{4}, 0 \right) \\
\notag &\phi_*^3= \left( \frac{1}{4} \left(\sqrt{-2m} -2 -\sqrt{4
-2m -20 \sqrt{-2m}}\right), \frac{1}{4} \left(-3\sqrt{-2m} -2
+\sqrt{4 -2m -20 \sqrt{-2m}}\right), 2 +\frac{\sqrt{-2m}}{4}, 0
\right)\\ \notag &\phi_*^4 =\left( \frac{1}{4} \left(\sqrt{-2m} -2
+\sqrt{4 -2m -20 \sqrt{-2m}}\right), \frac{1}{4}
\left(-3\sqrt{-2m} -2 -\sqrt{4 -2m -20 \sqrt{-2m}}\right), 2
+\frac{\sqrt{-m}}{4}, 0 \right)\\ \notag &\phi_*^5= \left(
\sqrt{-2m}, \frac{1}{4} \left(\sqrt{-2m}-4\right),
2-\frac{\sqrt{-2m}}{4}, -\frac{3}{2} \sqrt{\frac{-m}{2}} \right)\\
\notag &\phi_*^6= \left( -\sqrt{-2m}, \dfrac{1}{2}\left(-4
-\sqrt{-2m}\right), 2 +\frac{\sqrt{-2m}}{4},\frac{3}{2}
\sqrt{\frac{-m}{2}} \right) \, .
\end{align}
It can easily be inferred that the third and sixth equilibrium
points are the new ones derived from the inclusion of $\alpha H^4$
in Eq. (\ref{motion1}), while the rest coincide with the
equilibrium points of the pure vacuum $f(R)$ case
\cite{Odintsov:2017tbc}. Even though the points are the same, it
is worth inspecting their stability once again, simply because the
overall system has now been altered. In particular, the type of
stability may not be changed but in order to ascertain this, we
must inspect such a possibility. In the following, we shall
examine the stability of each equilibrium point. Before doing so,
we should note that all six equilibria fulfil the Friedmann
constraint of Eq. (\ref{motion3}), hence they are all viable
cosmological solutions of the Friedmann equation.

In order to make a qualitative analysis, we shall make use of the
Hartman-Grobman theorem, where basically the stability of a
hyperbolic equilibrium point of a non-linear system is
topologically equivalent to the stability of the linearized system
in the neighborhood of the particular equilibrium point.
Consequently, the (nonzero) eigenvalues of the Jacobian can
account for the stability \footnote{Given that an eigenvalue is
zero, the equilibrium point is no longer hyperbolic, thus the
Hartman-Grobman theorem cannot be applied \textit{per se}. In this
case, one must identify any centre manifold of the phase space and
reduce the system to the remaining stable or unstable manifolds,
where the Hartman-Grobman theorem applies normally.}. From its
definition, the Jacobian is given by the expression
$J=\sum_{i,j}\frac{\partial f_i}{\partial x_j}$, where in the case
at hand we have $f_i=\dfrac{\mathrm{d}x_i}{\mathrm{dN}}$. In
particular, the Jacobian has the following form,
\begin{equation}
\centering
\label{Jacobian}
J=\left(
\begin{array}{cccc}
 2 x^*_1-x^*_3+3 & 0 & 2 (1-2 x^*_4)-x_1 & -4 (x^*_3-2) \\
 x_2^* & x^*_1-2 x^*_3+4 & -4 (x^*_2+2) & 0 \\
 0 & 0 & -4 (x^*_3-2) & 0 \\
 x^*_4 & 0 & 2 x^*_4 & x^*_1+2 x^*_3-4 \\
\end{array}
\right)
\end{equation}
The Jacobian seems to be independent of the value of $m$, though
it depends on the fixed points, and the fixed points can
significantly be altered as $m$ changes. Hence, both the fixed
points and the Jacobian change with respect to the free parameter
$m$, indicating changes in the stability as well.

We are bound to omit $\phi_*^3$ and $\phi_*^4$, as they are
complex for all $m \leq 0$, hence they have no physical interest
as fixed points. Regarding the remaining four, we may reach the
following conclusions: \vspace{5mm}
\begin{table}[h!]
  \begin{center}
    \caption{\emph{\textbf{Equilibrium Points for the vacuum $f(R)$ autonomous system}}}
    \label{table1}
    \begin{tabular}{ |p{2cm}||p{6.9cm}|p{3cm}|p{4cm}|  }
     \hline
      \textbf{ Fixed Point} & \textbf{Eigenvalues} & \textbf{Characterization} & \textbf{Centre Manifolds}  \\
           \hline
     $\phi_*^1$ & $\frac{1}{4} \left(-5\sqrt{-2m} -2 -\sqrt{4 -2m +20 \sqrt{-2m}}\right)$, $\frac{1}{4} \left(3\sqrt{-2m} -2 -\sqrt{4 -2m +20\sqrt{-2m}}\right)$, $-\sqrt{1 +5\sqrt{-2m} -\frac{m}{2}}$ and $2\sqrt{-2m}$ & Saddle & For $m=0$ and $m=-8$
      \\  \hline
      $\phi_*^2$ & $\frac{1}{4} \left(-5\sqrt{-2m} -2 +\sqrt{4 -2m +20 \sqrt{-2m}}\right)$, $\frac{1}{4} \left(3\sqrt{-2m} +\sqrt{-2m -2 +20 \sqrt{-2m}}\right)$, $\sqrt{1 +5\sqrt{-2m} -\frac{m}{2}}$ and $2\sqrt{-2m}$& Saddle & For $m=0$
      \\  \hline
      $\phi_*^5$ & $\left( \right)$ $\frac{1}{4} \left(5\sqrt{-2m} +2 -\sqrt{4 -2m +20 \sqrt{-2m}}\right)$, $\frac{1}{4} \left(5\sqrt{-2m} +2 +\sqrt{-2m +2 +20 \sqrt{-2m}}\right)$, $2 \sqrt{-2m}$ and $2 \sqrt{-2m}$ & Unstable Node & For $m=0$
      \\  \hline
     $\phi_*^6$ & $\frac{1}{4} \left(2 -5\sqrt{-2m}-\sqrt{4 -2m -20 \sqrt{-2m}}\right)$, $\frac{1}{4} \left(-5\sqrt{-2m} +2 +\sqrt{4 -2m -20 \sqrt{-2m}}\right)$, $-2\sqrt{-2m}$ and $-2\sqrt{-2m}$ & Stable Node-Focus & For $m=0$
      \\ \hline
    \end{tabular}
  \end{center}
\end{table}
What we can see from Table \ref{table1} is that all equilibrium
points yield at least one centre manifold, usually for $m=0$,
which recall describes the quasi-de Sitter or de Sitter
cosmological eras. Only in two cases does the centre manifold
persist, in which cases a purely imaginary eigenvalue appears; it
is easy to see, though, that this imaginary eigenvalue is not
sufficient to alter the asymptotic stability of the particular
fixed points, on the contrary it consolidates it. Furthermore, we
observe that, in spite of bifurcations due to the constant values
of the parameter $m$, the local stability of the equilibria is not
affected. Consequently, different values of $m$ lead to a
translocation of the equilibria in the phase space, but do not
alter their stability. Finally, we see that $\phi_*^6$ is
characterized by local asymptotic stability, hence it is expected
to attract solutions and stand for a viable cosmological model. It
is important to note that interestingly, this equilibrium
$\phi_*^6$ is among the new ones, that have no counterpart in the
pure vacuum $f(R)$ case.

Since one cannot draw any more conclusions from this general
standpoint, we proceed by studying qualitatively and
quantitatively (numerically) the subspaces of the phase space
corresponding  to the quasi-de Sitter or de Sitter ($m=0$), the
matter dominated era ($m=-\dfrac{9}{2}$) and the radiation
dominated era ($m=-8$) cases. Before that, however, it is worth
mentioning that the system is partly integrable, noticing that Eq.
(\ref{dx3}) is independent of the other variables and integrable
itself, as well as the fact that $x_{1}$ is bound to follow
$x_{3}$ in specific cases.

\subsection{The integrability of $x_{3}$}

\begin{figure}[h!] 
\centering
\includegraphics[width=0.6\linewidth]{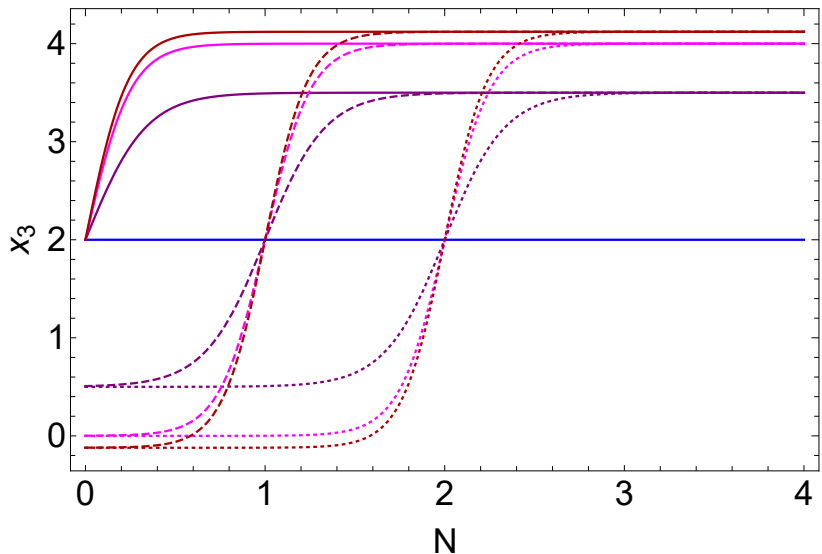}
\caption{Analytic solutions of Eq. (\ref{dx3}). Blue curves stand
for $m=0$, purple curves for $m = -\dfrac{9}{2}$ and red ones for
$m = -8$; the solid curves denote initial conditions for $N_{0} =
0$, while dashing becomes thinner as $N_{0} >0$ grows.}
\label{fig:x3_integrable}
\end{figure}

Eq. (\ref{dx3}) is easily solved and the solution is,
\begin{equation}\label{x3sol}
x_{3} (N) = \frac{1}{2} \left(4-\sqrt{2m} \tan \left(\sqrt{2m} (N-N_{0})\right)\right) \, ,
\end{equation}
where $N_{0}$ stands for the initial $e$-foldings value
(specifying the initial condition for $x_{3}$). Given specific
values for $m$ and $N_{0}$, we see in Fig.
(\ref{fig:x3_integrable}) that $x_{3}$ always converges -within at
most four $e$-foldings, to an equilibrium point given by,
\begin{equation*}
x_3^* = 2 +\frac{\sqrt{-2m}}{2} \, .
\end{equation*}
The blue curves in Fig. (\ref{fig:x3_integrable}) stand for $m =
0$ (de Sitter expansion), the deep purple curves for $m =
-\dfrac{9}{2}$ (matter-dominated cosmologies), the magenta curves
for $m = -8$ (radiation-dominated cosmologies) and the red curves
for $m = -9$ (stiff matter-dominated cosmologies), while solid,
dashed and dotted curves depict increasing $N_0$; all cases
exhibit the same behavior.
Interestingly, when $x_{3} = 2$, which is true only in the (quasi)
de Sitter expansion ($m = 0$), the system of Eqs. (\ref{dx1}),
(\ref{dx2}) and (\ref{dx4}) is significantly simplified, becoming
integrable as well. The solutions are,
\begin{equation}
x_1 (N) = \dfrac{ x_{1}(0) e^N }{1 + x_{1}(0) \left( 1-e^N \right)} \, , \;\;\;
x_2 (N) = \dfrac{ x_{2}(0) }{1 + x_{1}(0) \left( 1-e^N \right)} \;\;\; \text{and} \;\;\;
x_4 (N) = \dfrac{ x_{4}(0) }{1 + x_{1}(0) \left( 1-e^N \right)} \, ,
\end{equation}
where $x_{1}(0)$, $x_{2}(0)$ and $x_{4}(0)$ are arbitrarily chosen
initial values; all solutions are naturally lead to the phase
space points $x_{1}^* = -1$, $x_{2}^* = x_{4}^* = 0$. The
respective parametric solution is,
\begin{equation} \label{integrals}
x_{2} = x_{2} (0) \left( 1 + x_{1} \right) \;\;\; \text{and} \;\;\; x_{4} = x_{4}(0) \left( 1 + x_{1} \right) \, ,
\end{equation}
that define integrals of motion in the hyperplane $x_{3} = 2$ for
$m=0$. An important comment here is that the phase space point
$x_{1}^* = -1$, $x_{2}^* = x_{4}^* = 0$ and $x_{3} = 2$, was found
to be a stable fixed point of the pure $f(R)$ gravity case
\cite{Odintsov:2017tbc}. In the next section we further highlight
this important issue.

\subsection{The case of de Sitter Expansion: $m=0$}

Let us now proceed with the phase space behavior focusing on the
subspace related to de Sitter or quasi-de Sitter solutions, which
recall correspond to the specific value of $m$, that of $m=0$.
From the realization that $x_3=2$ when $m=0$, we obtain that the
deceleration parameter is exactly $q=-1$. Therefore, this choice
is suitable for describing a de-Sitter (or quasi-de Sitter)
expansion, that can describe both the inflationary and the
late-time accelerating expansions. These two different eras can be
described appropriately by choosing the final value of the
$e$-foldings number to be large or small, describing the late or
early time evolution of the Universe respectively. Also the
initial value of the $e$-foldings number may play some role
eventually, as we discuss in a later section, when the Big Rip
singularity is considered.

For $m=0$, the equilibrium points extracted from
Eq.(\ref{equilibrium}) are,
\begin{align}
\centering \label{phi*A}
\phi_*^1=&(-1,0,2,0)&\phi_*^2=&(0,-1,2,0)\, ,
\end{align}
where for simplicity we neglected the rest, since they merely
coincide with these two. The first yields the eigenvalues $\{-1,
-1, 0, -1\}$, where $\vec{v}_1 = (1,0,0,0)$, $\vec{v}_2 =
(0,1,0,0)$ and $\vec{v}_4 = (0,0,0,1)$ are tangent to the stable
manifolds and a centre manifold appears to be tangent on
$\vec{v}_{0} = (3,-4,1,0,0)$; seeing that the latter is removable
from the system due to the $\vec{v}_3$ direction, we can
characterize the equilibrium point as locally asymptotically
stable. The second fixed point exhibits the eigenvalues
$\{1,0,0,0\}$, where only $\vec{v}_{-1,1} = (-1,1,0,0)$ defines an
unstable manifold, and with all the rest being tangent to centre
manifolds, deeming the fixed point non-hyperbolic; as a result, we
are unable to characterize these by using the Hartman-Grobman
theorem.

Hence we easily come to the conclusion that when de Sitter or
quasi-de Sitter solutions are considered, which correspond to the
case $m=0$, the phase space substructure for the (quasi) de Sitter
solutions is exactly the same as in the pure $f(R)$ vacuum case
studied in Ref. \cite{Odintsov:2017tbc}, where the term $\alpha
H^4$ was absent. The solutions converge fast to the $x_{4}=0$
subspace, though numerical solutions seem to give an upper limit
to this convergence, revealing the destabilizing role of the
centre manifolds.

\begin{figure}[h!] 
\centering
\includegraphics[width=0.4\linewidth]{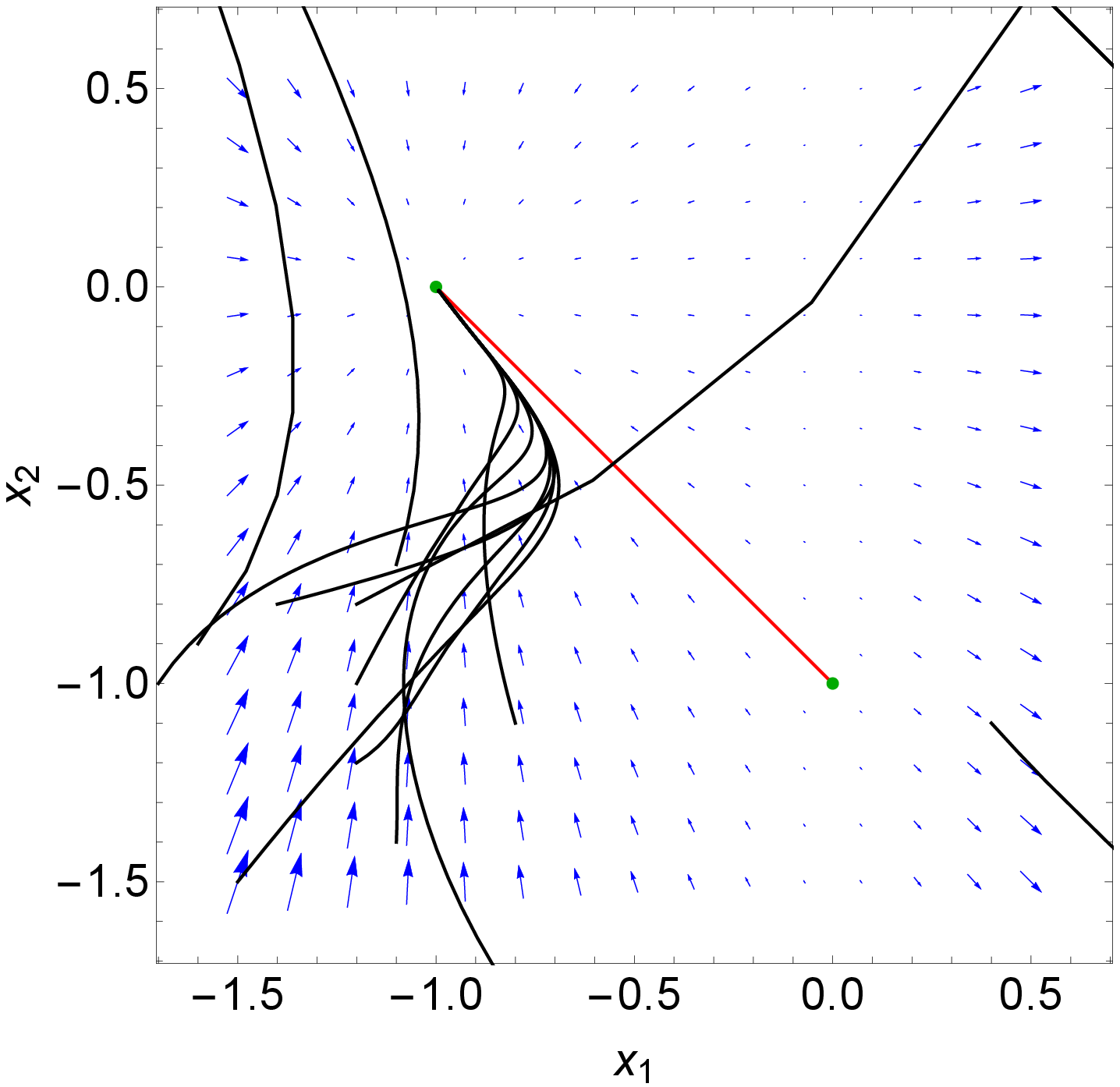}
\includegraphics[width=0.4\linewidth]{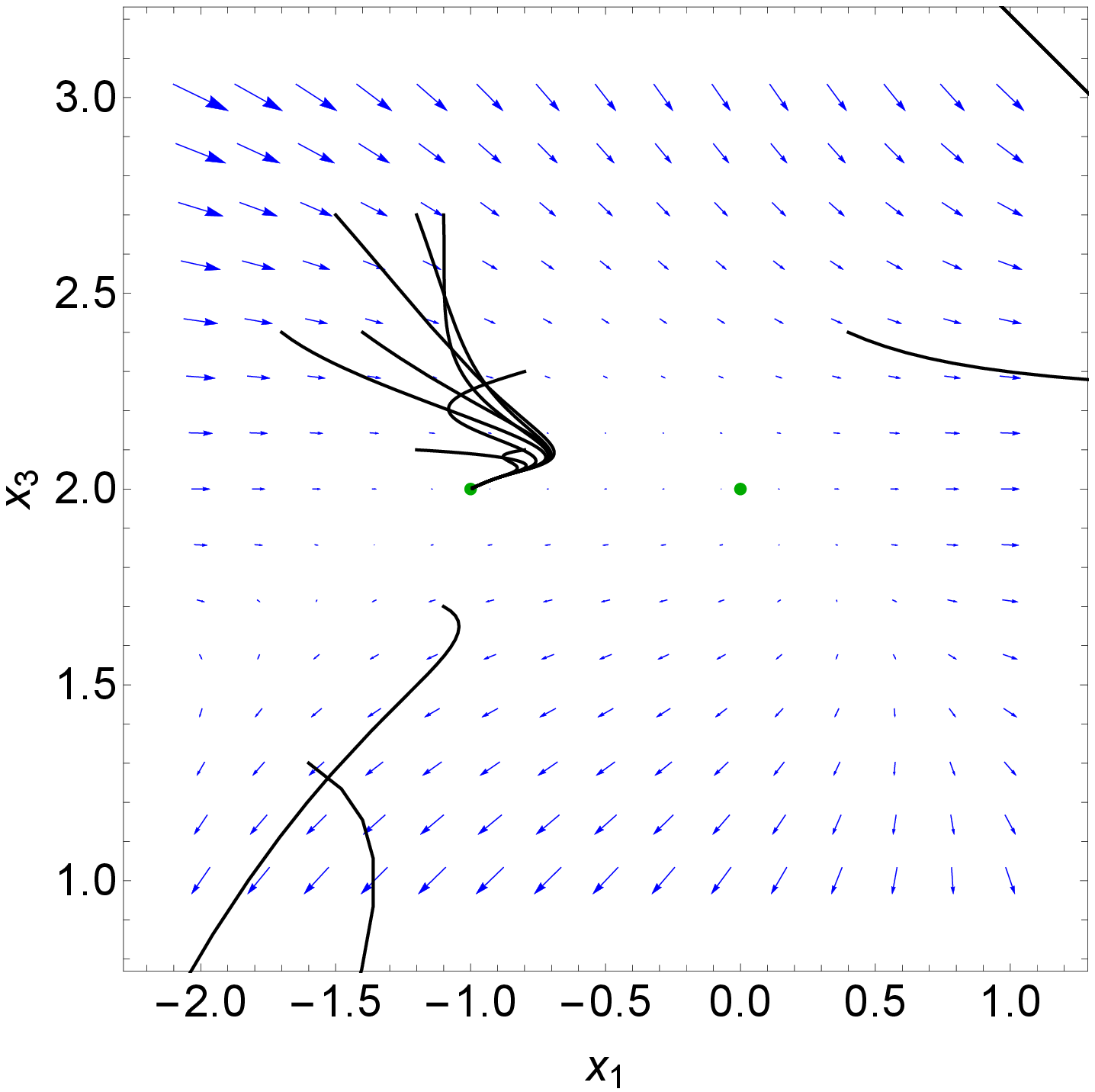}
\includegraphics[width=0.5\linewidth]{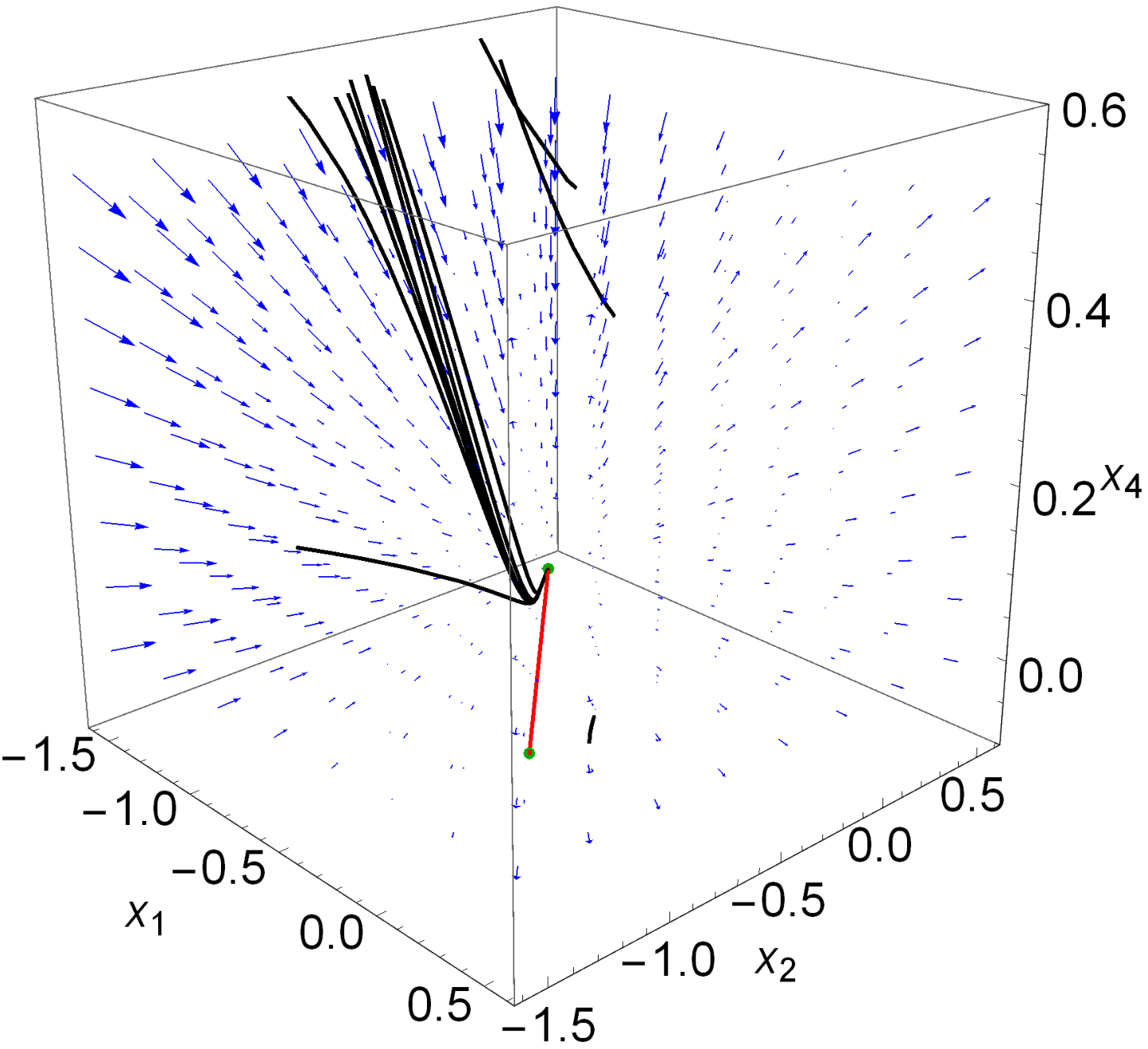}
\caption{The $x_1 - x_2$, $x_1 - x_3$ and $x_1 - x_2 - x_4$
subspaces of the phase space of the system of Eqs.
(\ref{dx1}-\ref{dx4}) for $m=0$. Blue vectors stand for the vector
field, black curves for the solutions, green dots for the
equilibrium points and the red line for the heteroclinc orbit.}
\label{fig:first_deSitter}
\end{figure}

An interesting scenario occurs when the initial conditions are
$x_{1}, x_{2}
> 0$ and/or $x_{3} < 2$, in which case the trajectory in the phase space for evolving $N$ is bound to diverge and tend to
infinity. The initial conditions that respect these boundaries
(hence for which $x_{1}, x_{2} < 0$ and $x_{3} > 2$ - also $x_{4}
> 0$ so that we are consistent) converge rapidly to equilibrium
$\phi_*^1$. The numerical solutions are depicted in Fig.
\ref{fig:first_deSitter}, where the vector field corresponds to
blue arrows, the numerical solutions as black curves and the
equilibria are green dots. The first subplot depicts the $x_{1} -
x_{2}$ slice of the phase space, the second subplot the $x_{1} -
x_{3}$ slice, and finally the third the $x_{1} - x_{2} - x_{4}$
hyperplane for $x_{3}^* = 2$.
Another interesting situation occurs, by using the integrals
defined by Eq. (\ref{integrals}) and setting $x_{2}(0) = -1$ and
$x_{4}(0) = 0$, to match the unstable equilibrium point
$\phi_*^2$. In this way, we obtain a heteroclinic trajectory
towards the stable equilibrium point $\phi_*^1$, that functions as
the unstable manifold of the former meets the stable manifold of
the latter. This is marked with a straight red line in Fig.
\ref{fig:first_deSitter}. This behavior is quite interesting since
it shows that there is a trajectory in the (quasi) de Sitter
subspace of the total phase space that actually connects the
unstable fixed point $\phi_*^2$ with the stable fixed point
$\phi_*^1$.

As we mentioned, only those trajectories whose initial values
respect the rule $x_{1}, x_{2} < 0$ , $x_{3} > 2$ and $x_{4} > 0$
converge to the $\phi_*^1$, while all the other trajectories are
led away by the instability of $\phi_*^2$. This is the outcome of
the centre manifolds dominating the neighborhood of $\phi_*^2$. In
fact, a higher-order approximation would deem it globally
unstable.

These results conclude the examination of the (quasi) de Sitter
solutions subspace of the total phase space. Our results indicate
that the thermal terms do not affect significantly the de Sitter
subspace of the total phase space, and thus, this means that no
thermal effects can be considered of having a measurable effect on
$f(R)$ gravity, even for late-time de Sitter solutions. To be
honest, thermal effects would make their presence noticeable near
a Big Rip singularity, not billion years before it. Thus our study
verified our initial suspicion that such thermal effects do not
affect the late-time era, at least when no Big Rip singularity is
present. This is the phase space picture though, but a focused
study on the late-time era observable quantities will show that
for quite large values of $\alpha$ appearing in
$\alpha\kappa^2H^4$, can actually alter the late-time behavior,
even if this late-time era occurs quite far away from the Big Rip
singularity, chronologically speaking.

Now in order to complete the phase space effects of the thermal
terms, we shall analyze the subspaces of the total phase space
corresponding to the matter and radiation dominated solutions.
This is the subject of the following two subsections.

\subsection{The case of Matter-Dominated Cosmologies: $m=-\dfrac{9}{2}$}

The subspace of the total phase space corresponding to
$m=-\dfrac{9}{2}$, describes the matter domination era, as we
discussed earlier. In this section we shall investigate
quantitatively the effects of the thermal term $\alpha\kappa^2H^4$
on the matter dominated solutions of the total phase space. In
this case, the equilibrium points are,
\begin{align}\centering
&\phi_*^1 =\left(
\frac{1}{4}\left(-5-\sqrt{73}\right),\frac{1}{4}\left(7+\sqrt{73}\right),\frac{1}{2},0
\right) \\ \notag &\phi_*^2 =\left(
\frac{1}{4}\left(-5+\sqrt{73}\right),\frac{1}{4}\left(7-\sqrt{73}\right),\frac{1}{2},0
\right) \\ \notag &\phi_*^5 =\left( 3, -\frac{1}{4}, \frac{1}{2},
-\frac{9}{4} \right) \\ \notag &\phi_*^6 =\left( -3, -\frac{7}{4},
\frac{7}{2},\frac{9}{4} \right) \, .
\end{align}
Here, two points are discarded, as they are complex and have no
physical meaning and no other equivalence is observed. It is
easily inferred that for $m=-\dfrac{9}{2}$, all fixed points are
hyperbolic since the real part of the respective eigenvalues is
nonzero.

More specifically, equilibria $\phi_*^1$ and $\phi^*_2$ yield the
eigenvalues $\{ -6.386,6,-4.272,-0.386001 \}$ and $\{
6,4.272,3.886,-2.114 \}$ respectively, being both characterized as
a saddles; interestingly, direction $\vec{v}_2 = (0,-1,0,0)$ is
always tangent to an unstable manifold. Equilibrium $\phi_*^5$
exhibits the eigenvalues $\{ 6.386,6,6,2.114 \}$, all being
positive, hence it is deemed an unstable node. Finally,
equilibrium $\phi_*^6$ is found to be a stable node-focus, as its
eigenvalues are $\{ -6,-6,-3.25+1.71391 i,-3.25-1.71391 i \}$, two
of them being real and negative also equal, hence a degeneracy is
present, and the other two being complex with negative real part;
solutions are expected to converge to the hyperplane parallel to
$\vec{v}_{-1,2,4} =(-1.5095, 0.3784, 0, 1)$ and then oscillate
towards the fourth equilibrium point.

This behavior is clearly observed in Fig. {\ref{fig:first_Matter},
where again the vector field is denoted as blue arrows, the
numerical solutions as black curves and the equilibria are green
dots. In the second subplot, where an $x_{1} - x_{3}$ slice is
shown, the solutions seem to converge simply on $x_{3} = 3.5$ and
then move towards the equilibrium, while in the first and third
subplots, that depict the subspace orthogonal to $x_{3}$
direction, we clearly see the oscillating pattern; unfortunately,
the convergence is too fast to see complete cyclical
behavior.\footnote{We should note that the only solution being
completely repelled has an initial value of $x_{3}(0) = 0.5$, so
it is caught by the unstable manifolds of $\phi_*^1$ and
$\phi_*^2$; yet, it is still repelled towards positive values of
$x_4$, in accordance to what we conclude. Furthermore, this again
implies that the phase space is split and yields viable
cosmological models only for $x_{3} > 0.5$.}
\begin{figure}[h!] 
\centering
\includegraphics[width=0.4\linewidth]{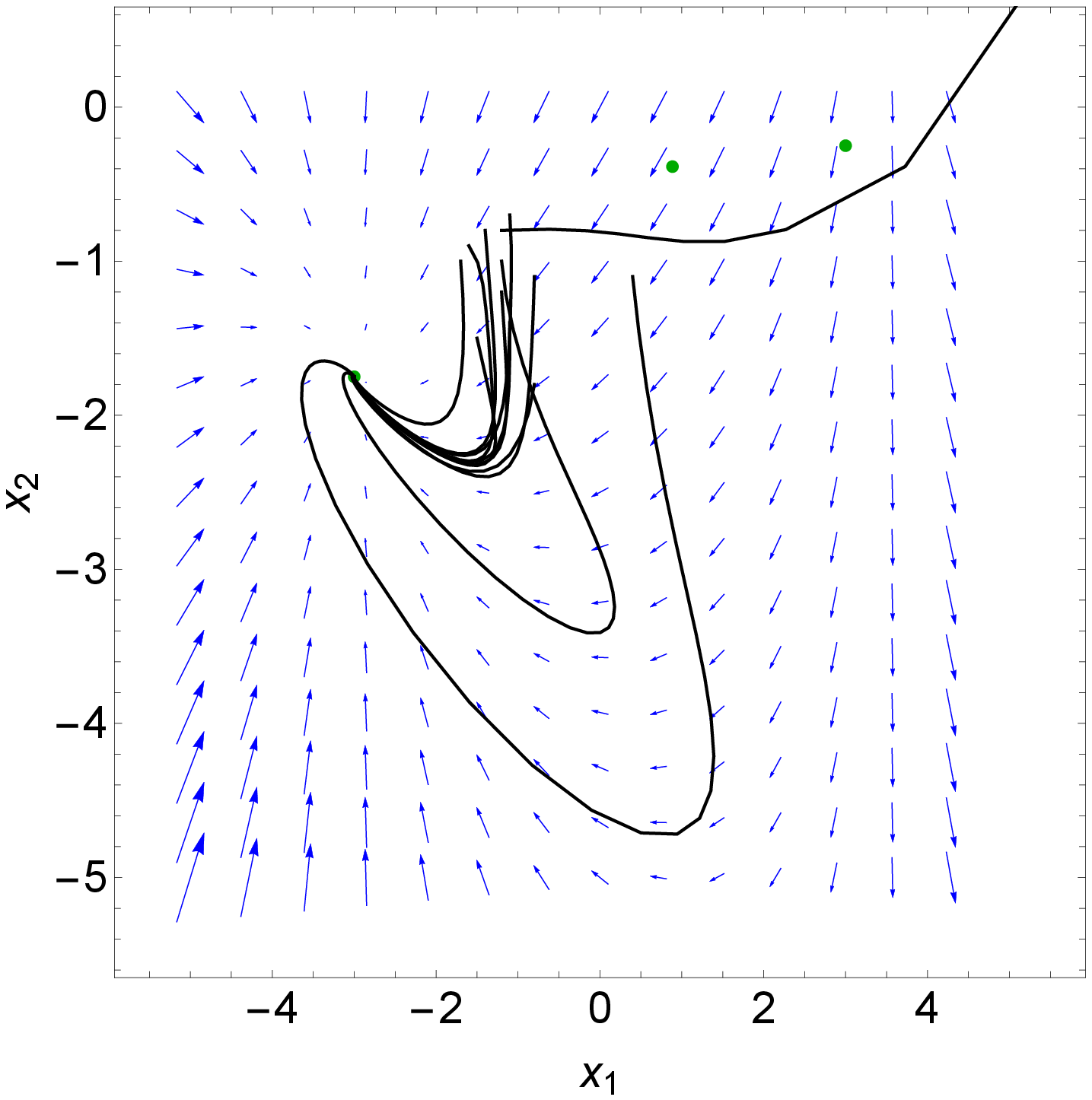}
\includegraphics[width=0.4\linewidth]{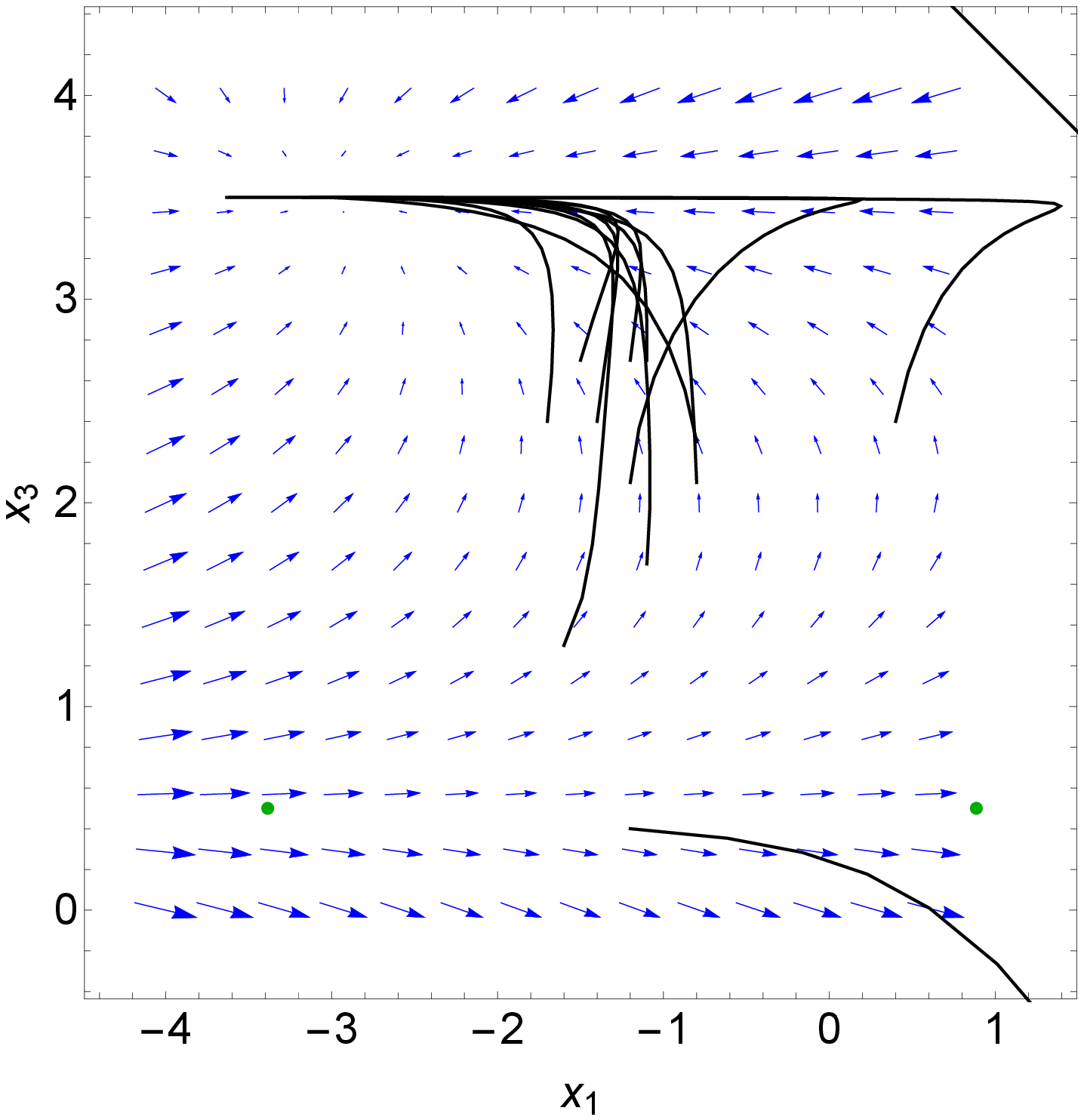}
\includegraphics[width=0.5\linewidth]{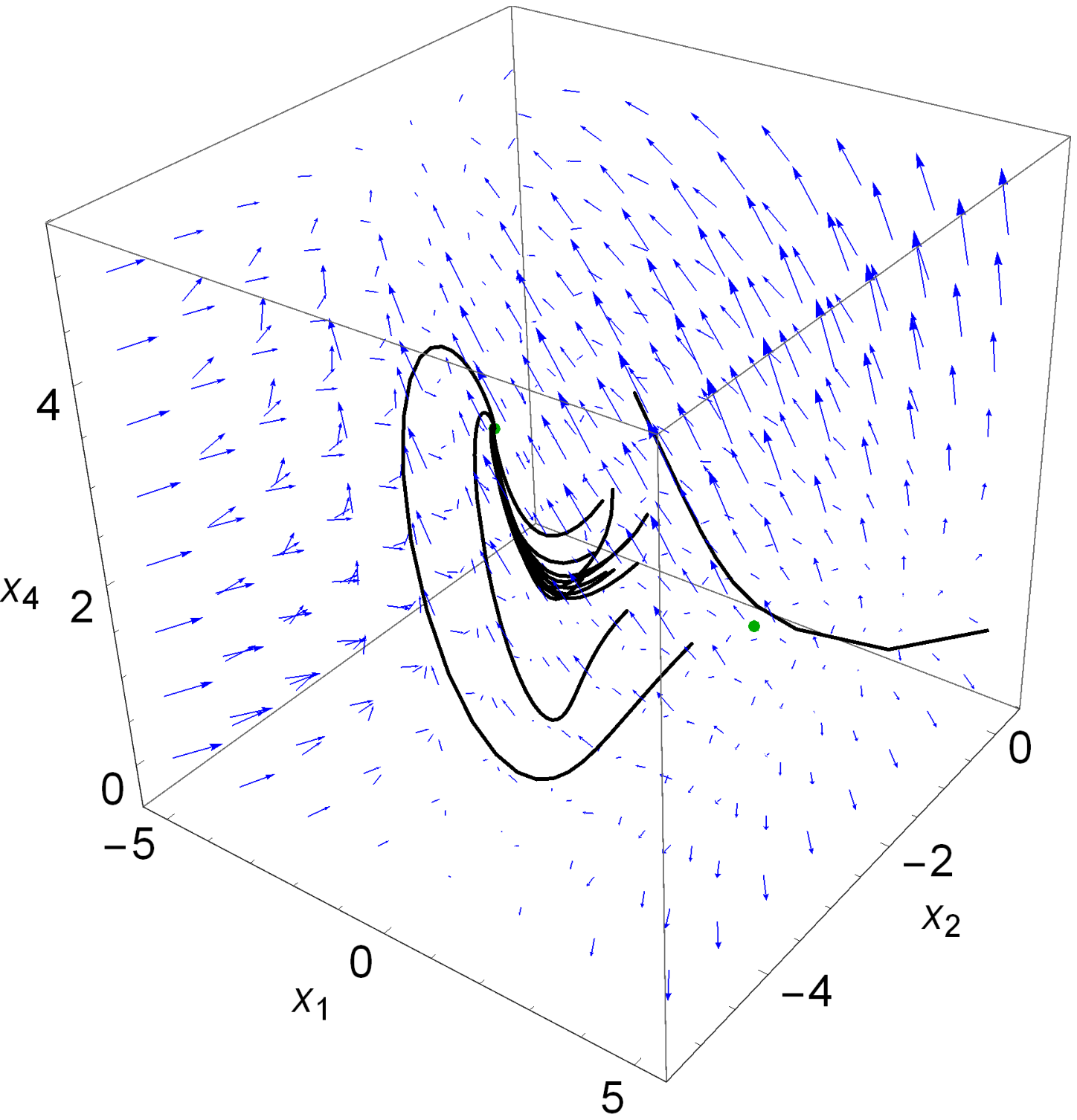}
\caption{The $x_1 - x_2$, $x_1 - x_3$ and $x_1 - x_2 - x_4$ subspaces of the phase space of the system of Eqs. (\ref{dx1}-\ref{dx4}) for $m = -\dfrac{9}{2}$. Blue vectors stand for the vector field, black curves for the solutions, green dots for the equilibrium points and the red line for the heteroclinc orbit.} \label{fig:first_Matter}
\end{figure}
Interestingly, the equilibrium point attained is the only one with
a viable (positive) nonzero value for $x_{4}$, while $\phi_*^1$
and $\phi_*^2$ that come along $x_{4} = 0$, which describes the
vacuum $f(R)$ theory. These are asymptotically unstable, repelling
solutions to $x_{4}>0$. Consequently, the additional thermal term
$\alpha\kappa^2H^4$ becomes important as we escape the
inflationary era or the late-time accelerating expansion and we
enter the more typical cases of matter dominated cosmologies. This
result is quite intriguing since it shows that when a de Sitter
era is followed by some matter dominated era, the thermal effects
term may affect actually the dynamical evolution of the system, in
the way we described in this subsection.

\subsection{The case of Radiation-Dominated Cosmologies: $m=-8$}

Now let us consider the effects of the thermal term
$\alpha\kappa^2H^4$ on the phase space structure corresponding to
radiation dominated solutions. The results for the
radiation-dominated era are similar to those obtained for the
matter-dominated solutions, as we now evince. Setting $m = -8$, a
value which specifies the subspace of the total phase space
corresponding to radiation dominated solutions, four equilibria
are obtained,
\begin{align}\centering
&\phi_*^1 =\left( -4,5,0,0 \right) \\ \notag &\phi_*^2 =\left(
1,0,0,0 \right) \\ \notag &\phi_*^5= \left( 4,0,0,-3 \right) \\
\notag &\phi_*^6 =\left( -4,-2,4,3 \right) \, .
\end{align}
In the same manner, two points are discarded, as they are complex
and have no physical meaning; all other fixed points are
hyperbolic, as the real part of the respective eigenvalues is
nonzero. As in the case of $m = -\dfrac{9}{2}$, the three first
equilibria are asymptotically unstable, whereas $\phi_*^6$ is
stable signifying a viable cosmological solution, which again
comes with a positive value of $x_4$, hence it differs
significantly from the vacuum $f(R)$ case. Specifically, the
eigenvalues for $\phi_*^1$ and $\phi_*^2$ are $\{ -8,8,-5,0 \}$
and $\{ 8,5,5,-3 \}$ respectively, deeming them saddles; again the
$\vec{v}_{2} = (0,1,0,0)$ is tangent to an unstable manifold for
the second, but a centre manifold exists for the first fixed
point. The fixed point $\phi_*^5$ yields the eigenvalues $\{
8,8,8,3 \}$, so this fixed point is a degenerate unstable node.
Finally, $\phi_*^6$ exhibits the eigenvalues $\{ -8,-8,-4.5 +
1.93649i, -4.5 - 1.93649i \}$, hence it is a stable node-focus;
the solutions are attracted by degenerate stable manifolds tangent
to $\vec{v}_{-3,4} = (0,0,-4,3)$ and $\vec{v}_{2}=(0,1,0,0)$ to
the hyperplane orthogonal to $\vec{v}_{-1,2,4} = (-1.5, 0.5, 0,
1)$, where they oscillate towards the fixed point. In Fig.
\ref{fig:first_Radiation} we present the phase space behavior of
several trajectories corresponding to radiation domination
solutions.
\begin{figure}[h!] 
\centering
\includegraphics[width=0.4\linewidth]{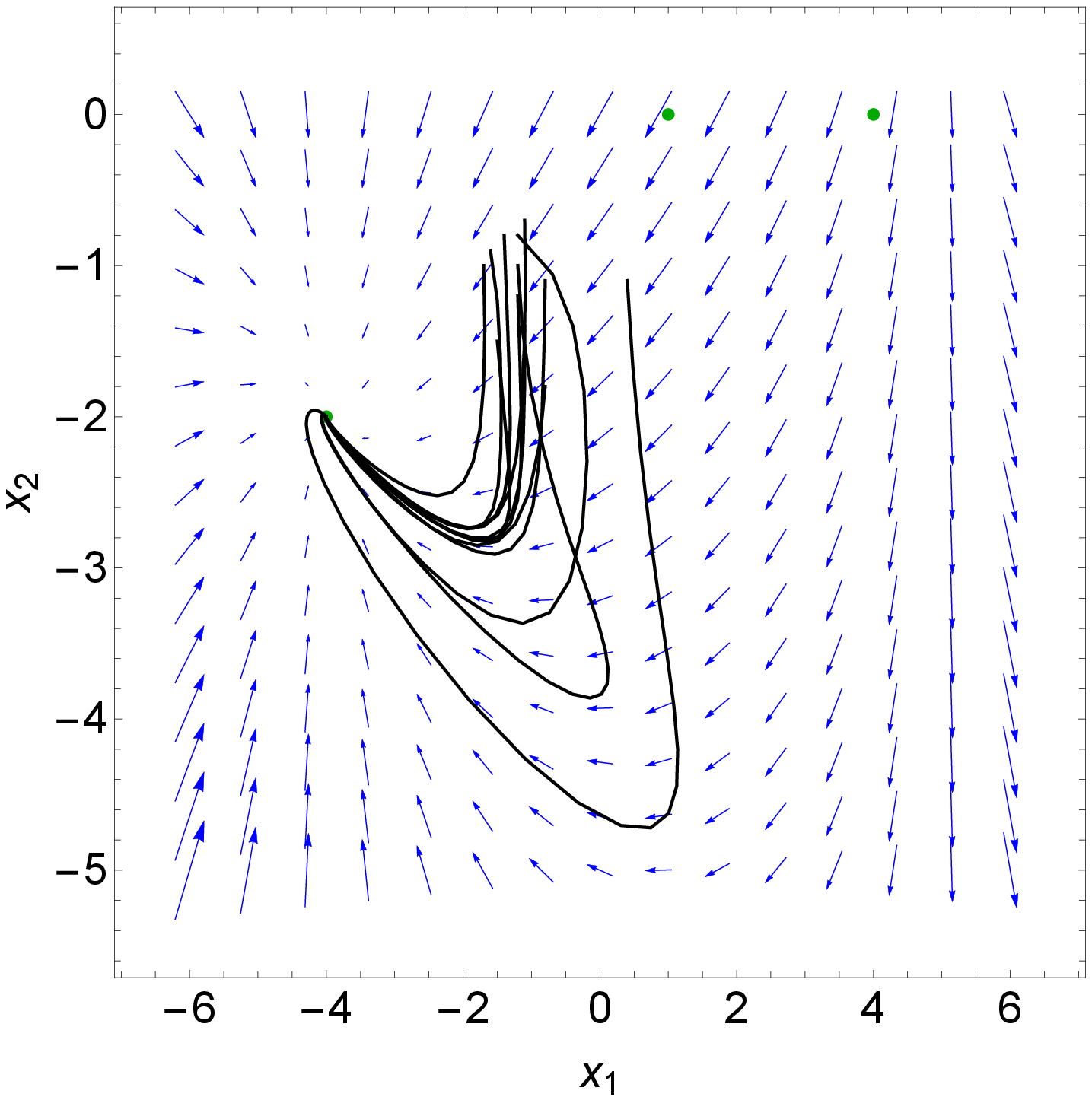}
\includegraphics[width=0.4\linewidth]{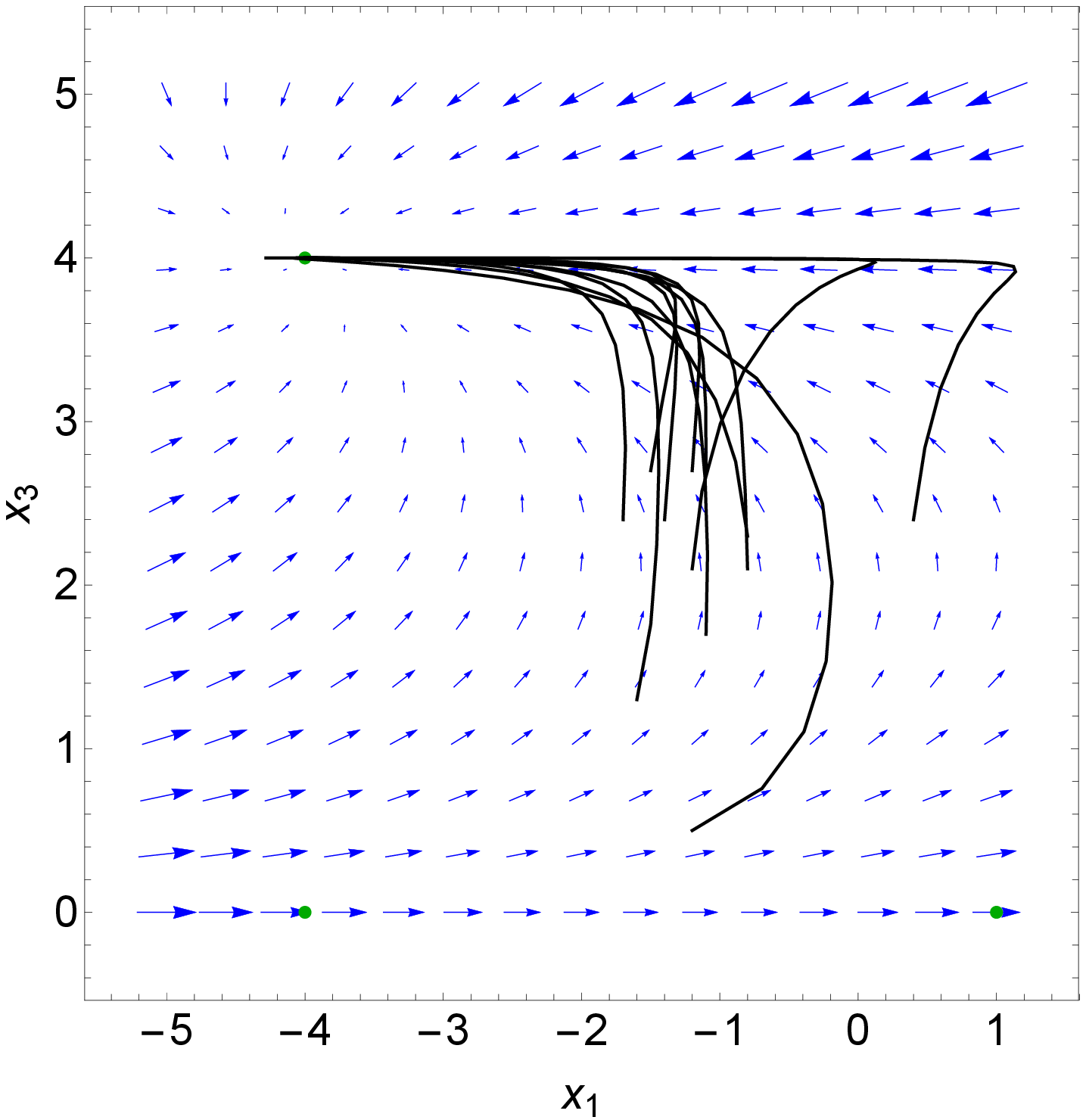}
\includegraphics[width=0.5\linewidth]{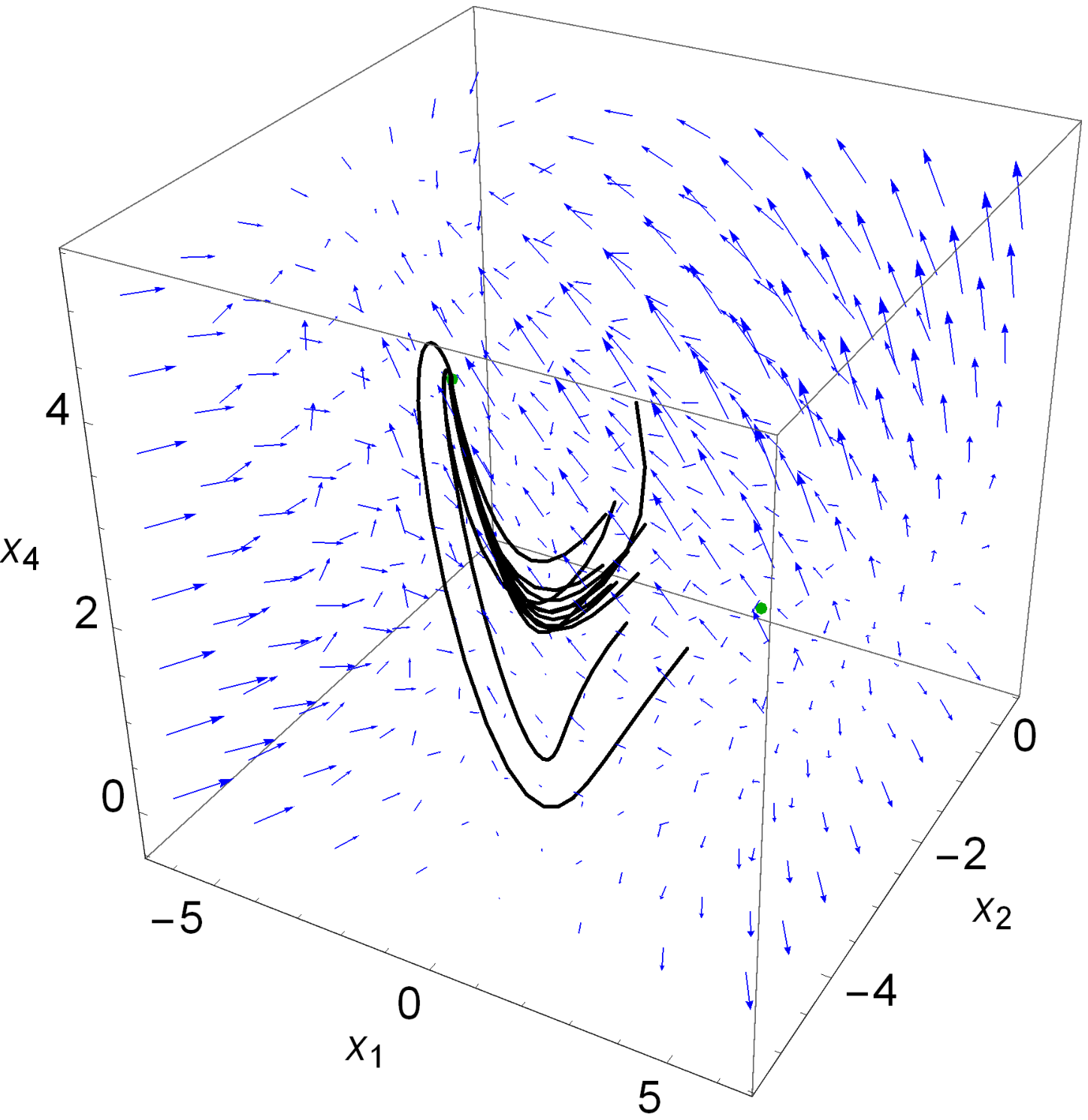}
\caption{The $x_1 - x_2$, $x_1 - x_3$ and $x_1 - x_2 - x_4$
subspaces of the phase space of the system of Eqs.
(\ref{dx1}-\ref{dx4}) for $m = -8$. Blue vectors stand for the
vector field, black curves for the solutions, green dots for the
equilibrium points and the red line for the heteroclinic orbit.}
\label{fig:first_Radiation}
\end{figure}
Our results indicate that the effects of the thermal terms on the
matter and radiation dominated solutions of the total subspace, is
significant, since the thermal term changes the phase space
structure of vacuum $f(R)$ gravity. This is somewhat surprising to
us, and also reportable result.

\section{The Case of the Vacuum $f(R)$ non-Autonomous System}

The cases we examined so far where those that rendered the
dynamical system of Eqs. (\ref{dx1})-(\ref{dx4}) autonomous, those
cases for which the Hubble rate had such a form as a function of
time, that parameter $m$ takes constant values. However, this does
not exhaust all interesting cases, especially the cases for which
a finite-time singularity is approached at some time $t_s$. In
this section we shall consider the Big Rip case, and we shall
investigate the effects of the thermal term $\alpha H^4$ on
the phase space of $f(R)$ gravity, as a Big Rip singularity is
approached.

Following Ref. \cite{Nojiri:2005sx}, we can assume that the Hubble
rate is given as
\begin{equation}
H(t) \simeq H_0 \left( t - t_s \right)^{-\beta} \, ,
\end{equation}
where $H_0$ some positive constant of dimension $sec^{\beta-1}$,
and $\beta$ a free parameter that allows for the classification
of the four types of singularities.

According to it, the Hubble rate reaches zero or infinity
-depending on the values of $\beta$- whenever $t = t_s$; at this
moment in time, some of the following singularities occur.
\begin{enumerate}

\item \underline{Type I Singularity (Big Rip)}, where $\beta >
1$: In this case, the Universe expands so rapidly due to its
growing content, that it is eventually ``ripped apart''. All the
scale factor, the total effective energy density and the total
effective pressure of the Universe diverge strongly as $\beta
\rightarrow \infty$, $\rho_{eff} \rightarrow \infty$ and $P_{eff}
\rightarrow \infty$.

\item \underline{Type II Singularity (Sudden Death)}, where $-1 <
\beta < 0$: In this case, the expansion of the Universe and the
energy density of its content reach constant values, $\beta
\rightarrow \beta_s$, $\rho_{eff} \rightarrow \rho_s$, however the
total effective pressure approaches infinity, $P_{eff} \rightarrow
\infty$.

\item \underline{Type III Singularity}, where $0 < \beta < 1$: In
this case, the Universe reaches a constant rate of expansion, but
its actual or effective content continues to grow exponentially.
Hence, while the scale factor reaches a constant value, $\beta
\rightarrow \beta_s$, the total effective energy density and the total
effective pressure blow up, $\rho_{eff} \rightarrow \infty$ and
$P_{eff} \rightarrow \infty$.

\item \underline{Type IV Singularity}, where $\beta < -1$: In
this final case, the Universe and its content grow until a
specific point, hence the scale factor, the total effective energy
density and the total effective pressure simply reach constant
values, $\beta \rightarrow \beta_s$, $\rho_{eff} \rightarrow \rho_s$ and
$P_{eff} \rightarrow P_s$. However, the higher-than-second
derivatives of the Hubble rate diverge. This is the only case of
the finite-time singularities that a Universe may reach and
smoothly pass through without being actually affected, or at least
without being violently affected.
\end{enumerate}

Rewriting the Hubble rate with respect to the $e$-foldings number,
we have,
\begin{equation}\label{finiterate}
H(N) \simeq ( 1-\beta ) \left( N - N_c \right)^{\frac{\beta}{\beta-1}} \, ,
\end{equation}
where $N_c$ is an integration constant, depending on the initial
conditions. As a result, our parameter is no longer constant and
depends on the $e$-foldings number, taking the form
\begin{equation}
m = -\dfrac{\beta ( \beta+1 )}{ (\beta - 1)^2 \left( N_c - N \right)^2 } \, .
\end{equation}

Consequently, the system is no longer autonomous. Eqs.
(\ref{dx1})-(\ref{dx4}) now take the form,
\begin{align}\label{finitesystem}
\frac{\mathrm{d}x_1}{\mathrm{dN}} &= x_1(x_1+3-x_3) +2(x_3-2)(1-2x_4) \, , \\ \notag
\frac{\mathrm{d}x_2}{\mathrm{dN}} &= x_2(x_1-2x_3+4) -4(x_3-2) -\dfrac{\beta ( \beta+1 )}{ (\beta - 1)^2 \left( N_c - N \right)^2 } \, , \\ \notag
\frac{\mathrm{d}x_3}{\mathrm{dN}} &= \dfrac{\beta ( \beta+1 )}{ (\beta - 1)^2 \left( N_c - N \right)^2 } -2(x_3-2)^2 \;\;\; \text{and} \\ \notag
\frac{\mathrm{d}x_4}{\mathrm{dN}} &= x_4(x_1+2x_3-4) \, .
\end{align}
This dynamical system cannot be analyzed qualitatively as we did
so far, due to its ``time-dependent'' terms; it does not contain
any fixed points or other invariant sets \textit{per se}, but only
as a special case, hence we may only proceed to either analytical
or numerical solutions.

\subsection{The ``Big Rip'' singularity}

Depending on the nature of the finite-time singularity, we may
simplify the parameter $m$. Concerning the Type I (Big Rip)
singularity, we can utilize the approximation,
\begin{equation}
m \simeq -\dfrac{1}{\left( N_c - N \right)^2} \, .
\end{equation}
In this case, $m$ approaches zero, as $N$ tends to infinity, which
actually quantifies the effect of reaching the singularity. Simply
stated, as $N\to \infty$, the Big Rip singularity is approached.
The resulting dynamical system is identical to the one
corresponding to the (quasi) de-Sitter study we performed in an
earlier subsection. The value of $N_c$ itself plays an important
role, as it defines the route by which the de Sitter phase will be
reached, and thus whether it will be stable as we concluded above.

We can again see that Eq. (\ref{dx3}) is analytically integrated,
yielding in first order approximation close to $N_c$,
\begin{equation}
x_3 (N) \simeq 2 - \dfrac{1}{2(N -N_c)} \, ,
\end{equation}
which approaches minus infinity as $N \rightarrow N_c$ and $2$ as
$N \rightarrow \infty$. Utilizing this approximation of $x_3$
close to $N_c$, we can analytically integrate the complete system
as
\begin{equation}
\begin{split}
x_1 (N) &= 2+\dfrac{3 e^{N_c} \left(2 \sqrt{N_c} \mathcal{D}\left(\sqrt{N_c}\right) (2 N_c(x_1(0)+1) -3) -2 N_c (x_1(0)+1) -x_1(0) +2\right)}{\sqrt{-N_c} (2N_c (x_1(0)+1) -3)}\sqrt{N-N_c}  +\\
&+12 (N -N_c) +\dfrac{15 e^{N_c} \left(2 \sqrt{N_c} \mathcal{D}\left(\sqrt{N_c}\right) (2 N_c (x_1 (0)+1) -3) -2 N_c(x_1 (0) +1)-x_1(0) +2\right)}{\sqrt{-N_c} (2N_c (x_1(0)+1) -3)} (N -N_c)^{3/2} + \\
&+\left(40 -\frac{6 e^{2N_c} \left(-2 \sqrt{N_c} \mathcal{D}\left(\sqrt{N_c}\right) (2N_c (x_1(0)+1) -3) +2N_c (x_1(0)+1) +x_1(0) -2\right)^2}{N_c (3 -2N_c (x_1(0)+1))^2}\right) (N -N_c)^2  + \\
&+\dfrac{147 e^{N_c} \left(2 \sqrt{N_c} \mathcal{D}\left(\sqrt{N_c}\right) (2N_c (x_1(0)+1) -3) -2N_c (x_1(0)+1) -x_1(0) +2\right)}{2\sqrt{-N-c} (2N_c (x_1(0)+1) -3)} (N -N_c)^{5/2} + \\
&+\mathcal{O}\left((N -N_c)^3\right) \, ,
\end{split}
\end{equation}
\begin{equation}
\begin{split}
x_2 (N) &= \dfrac{1}{2 (N -N_c)} -3 +\dfrac{{3 e^N_c} \left(\frac{2 N_c (x_1(0)+1)+x_1(0) -2}{2N_c (x_1(0)+1)-3} -2 \sqrt{N_c} \mathcal{D}\left(\sqrt{N_c}\right)\right)}{\sqrt{-N_c}} \sqrt{N -N_c} + \\
&+\dfrac{\left(6N_c \left(-8N_c^2 (x_1(0)+1) +12N_c +x_1(0) +x_2(0) +1\right)+ 3\right)}{2N_c^2 (2N_c (x_1(0)+1) -3)} (N -N_c) - \\
&-\dfrac{15\left(e^{N_c} \left(2 \sqrt{N_c} \mathcal{D}\left(\sqrt{N_c}\right) (2N_c (x_1(0)+1) -3) -2N_c (x_1(0)+1) -x_1(0) +2\right)\right)}{\sqrt{-N_c} (2N_c (x_1(0)+1) -3)}  (N -N_c)^{3/2} + \\
&+\mathcal{O}\left((N -N_c)^2\right) \;\;\; \text{and}
\end{split}
\end{equation}
\begin{equation}
\begin{split}
x_3 (N) &= \dfrac{3 N_c x_4(0)}{(N -N_c) (2N_c (x_1(0)+1) -3)}+\dfrac{6 N_c x_4(0)}{2N_c (x_1(0)+1) -3} -\\
&-\dfrac{6 \left(e^{N_c} \sqrt{-N_c} x_4(0) \left(2 \sqrt{N_c} \mathcal{D}\left(\sqrt{N_c}\right) (2N_c (x_1(0)+1) -3) -2N_c (x_1(0)+1) -x_1(0) +2\right)\right)}{(3 -2N_c (x_1(0)+1))^2} \sqrt{N -N_c} + \\
&+\dfrac{24N_c x_4(0)}{2N_c (x_1(0)+1) -3} (N -N_c) - \\
&-\dfrac{30\left(e^{N_c} \sqrt{-N_c} x_4(0) \left(2 \sqrt{N_c} \mathcal{D}\left(\sqrt{N_c}\right) (2N_c (x_1(0)+1) -3) -2N_c (x_1(0)+1) -x_1(0) +2\right)\right)}{(3 -2N_c (x_1(0)+1))^2} (N -N_c)^{3/2} + \\
&+\mathcal{O}\left((N -N_c)^2\right) \, ,
\end{split}
\end{equation}
where $x_1(0)$, $x_2(0)$ and $x_4(0)$ are arbitrary initial
conditions and $\mathcal{D}(z) = \left( e^{-z^2}\right) \int_0^z
e^{y^2} \, dy$ is the Dawson integral function. As $x_3$ and $x_4$
approach $2$ and $0$ respectively, these solutions approach,
\begin{align*}
& x_1 (N) \simeq \dfrac{e^N x_1(0)}{x_1(0) \left(1 -e^N \right)+1} \, , \\ \notag
& x_2 (N) \simeq \dfrac{e^{N_c} N_c x_1(0) (N_c -N) \left( \mathrm{Ei}(N-N_c) -\mathrm{Ei}(-N_c) \right)+\left( e^N -1 \right) N_c x_1(0) +N_c x_2(0) (N_c -N) -N}{N_c (N -N_c) \left(\left(e^N -1\right) x_1(0) -1\right)} \;\;\; \text{and} \\ \notag
& x3 (N) \simeq \dfrac{x_4(0)}{x_1(0) \left(1 -e^N \right)+1} \, ,
\end{align*}
where $\mathrm{Ei}(z) = -\int_{-z}^{\infty } \dfrac{e^{-t}}{t} \, dt$ the exponential integral function.

As we stated, for very large $N$, $m$ tends to zero, hence the de
Sitter case should be reconstructed. However, this reconstruction
depends highly on the value of $N_c$. For all physical values $N_c
> 0$, so the initial conditions of the system are set prior to
$N_c$, the de Sitter case is not attained and the solutions
diverge as $N \rightarrow N_c$; as we can see in Fig.
\ref{fig:first_BigRip2}, the solutions are originally attracted by
the de Sitter stable fixed point, but then they are repelled, and
the trajectories in the phase space are attracted by an unstable
manifold.

This manifold can be located if we integrate analytically the
system of Eqs. (\ref{dx1}-\ref{dx3}) for $x_4 = 0$ and $x_3 = 2
-\dfrac{1}{(N -N_c)^2}$ and assuming that $N \rightarrow N_c$; the
manifold is indeed located as,
\begin{align}\label{bigripattractor}
\bar{x}_1 (N) &= \dfrac{N_c +2(1 -e^N) }{N_c +2 e^N (N -N_c -1) +2} \, , \\
\bar{x}_2 (N) &= \dfrac{5 e^{N -N_c} (N -N_c)(2 -N_c) \big( \mathrm{Ei}\left(N_c\right) - \mathrm{Ei}\left(N_c -N\right) \big) +5N_c -5N +1}{2e^N (2N_c -2N -2) - 2N_c -4} + \\
&+\dfrac{e^N \left(N^2 (1 -5N_c) +10N N_c (N_c +1) +8N_c (N -N_c)^2 \ln \big( \frac{N_c -N}{N_c} \big) -N_c (N_c +2) (5N_c +1)\right)}{2N_c (N -N_c) \left(2e^N (N -N_c -1) +N_c +2\right)} \;\;\; \text{and} \\
\bar{x}_3 (N) &= \frac{4N N_c -N -4N_c^2}{2N_c (N -N_c)} \, .
\end{align}
\begin{figure}[h!] 
\centering
\includegraphics[width=0.4\linewidth]{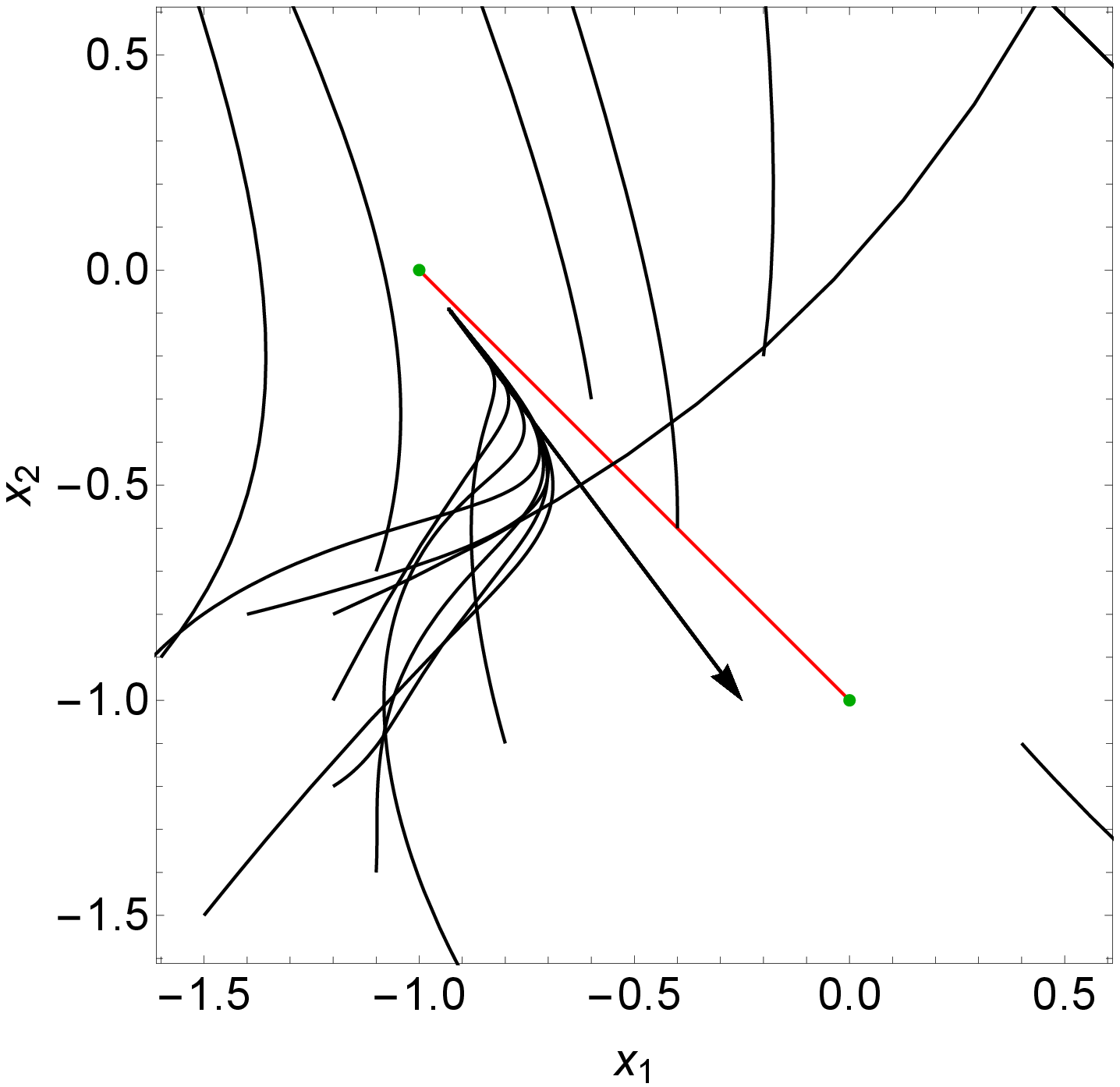}
\includegraphics[width=0.4\linewidth]{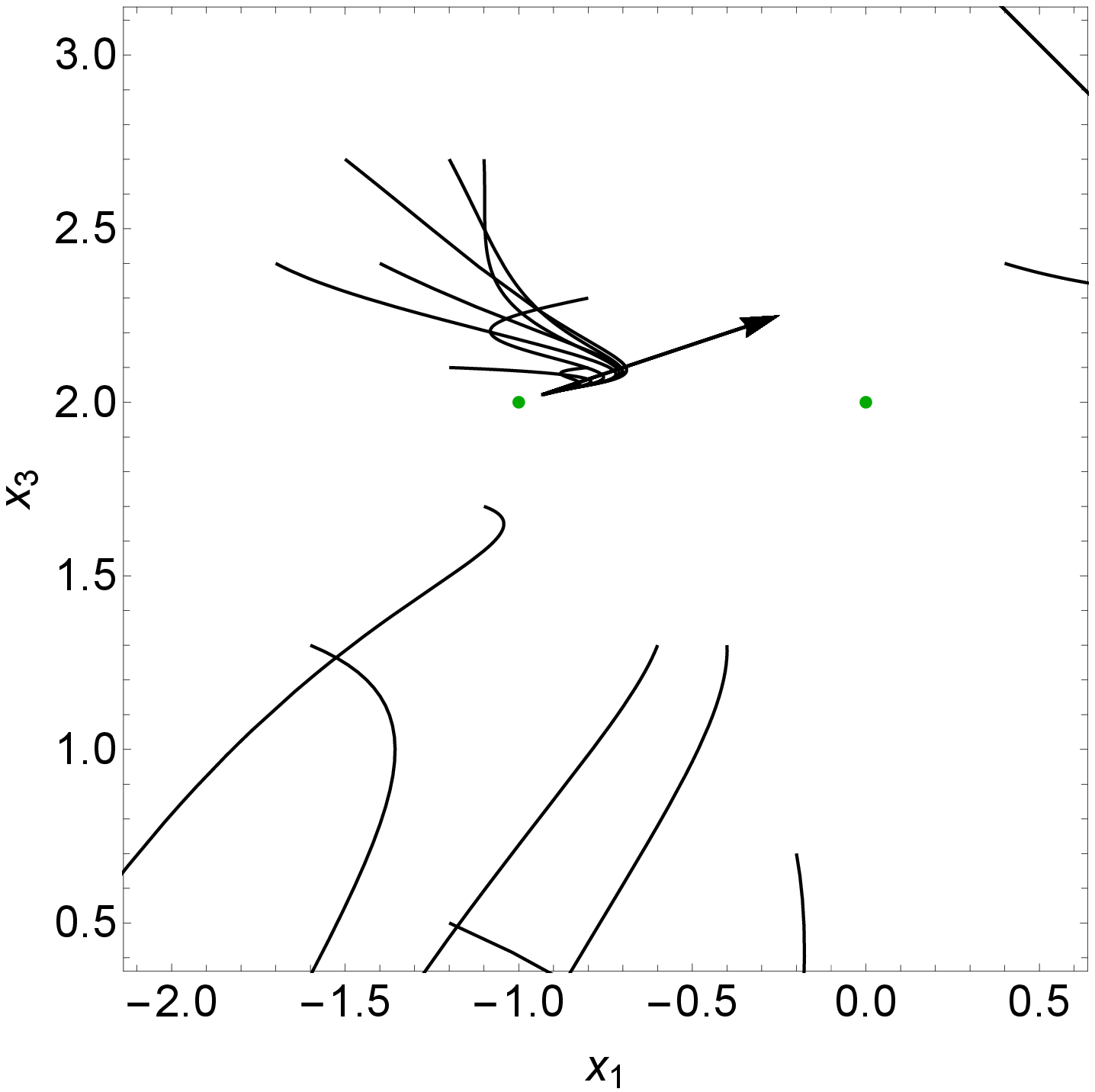}
\includegraphics[width=0.5\linewidth]{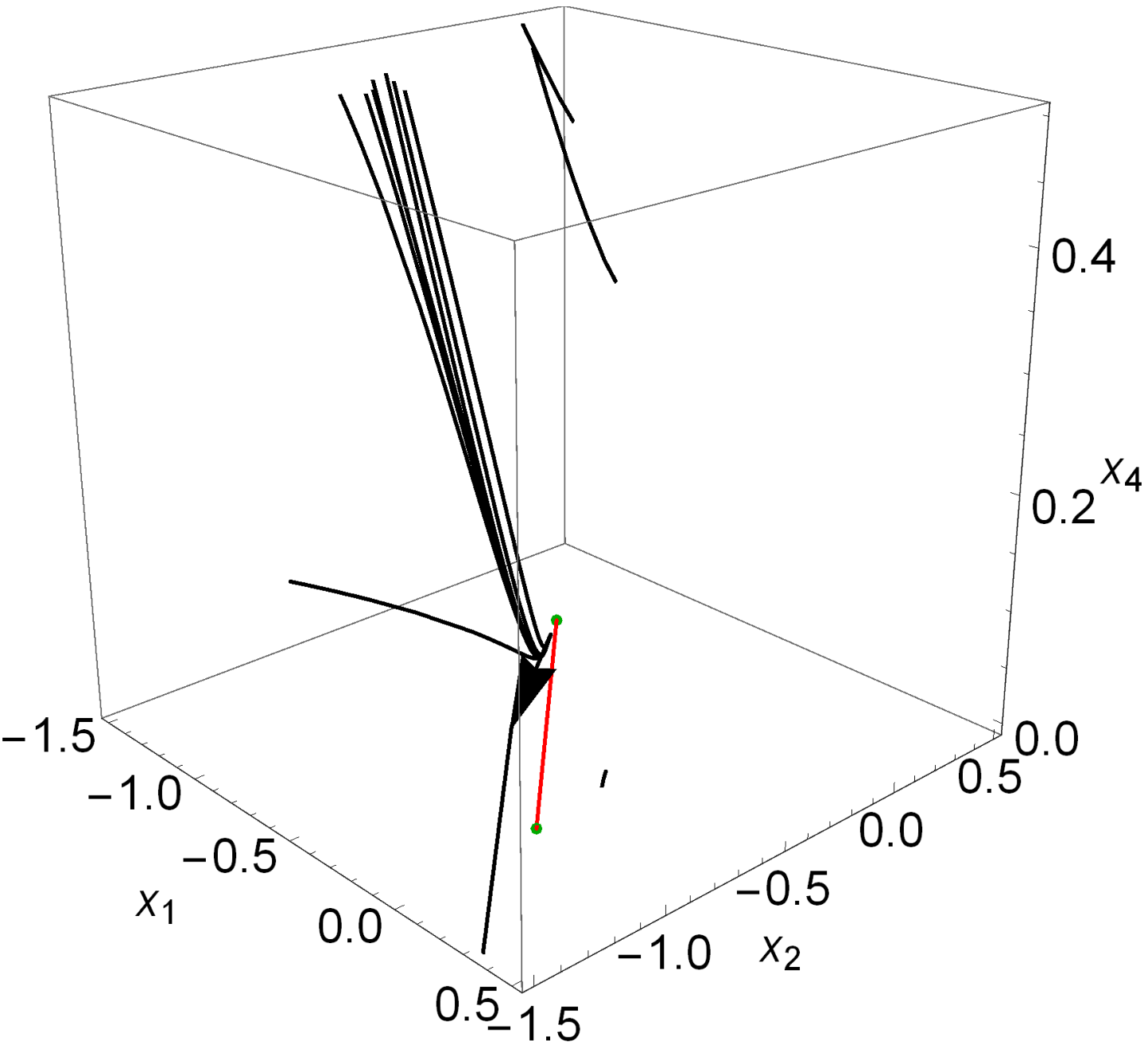}
\caption{ The $x_1 - x_2$, $x_1 - x_3$ and $x_1 - x_2 - x_4$
subspaces of the phase space of the system of Eqs.
(\ref{dx1}-\ref{dx4}) for $m = -\dfrac{1}{(N_c - N)^2}$ and $N_c >
0$. Black curves for the solutions, green dots for the equilibrium
points and the red line for the heteroclinc orbit.}
\label{fig:first_BigRip2}
\end{figure}

\section{Autonomous Dynamical System In the Presence of Perfect Fluids}

In the previous section we studied the vacuum $f(R)$ gravity phase
space subspaces corresponding to the (quasi) de Sitter, matter and
radiation dominated eras, discussing mainly the case that these
eras are generated by the vacuum $f(R)$ gravity. In this section
we shall generalize the theoretical framework, by taking into
account perfect fluids that can actually themselves can realize
the matter and radiation domination eras. Thus the study and
considerations of the previous sections are going to be
generalized to include the presence of perfect fluids.

So in the presence of perfect matter fluids, the gravitational
action reads,
\begin{equation}
\centering \label{action2}
S=\int{d^4x\sqrt{-g}\left(\frac{f(R)}{2\kappa^2}+\mathcal{L}_{eff}+\mathcal{L}_{matter}\right)}\,
,
\end{equation}
where $\mathcal{L}_{matter}$ is the Lagrangian corresponding to
perfect matter fluids. The inclusion of matter in the
gravitational action results in the emergence of the energy
densities of the perfect fluids in the equations of motion, with
the Friedmann equation becoming,
\begin{equation}
\centering
\label{motion4}
0=-\frac{f(R)}{2}+3F(\dot H+H^2)-18(4H^2\dot H+H\ddot H)\frac{dF}{dR}+\alpha\kappa^2H^2+\kappa^2\rho_{matter}\, ,
\end{equation}
where $\rho_{matter}=\rho_m + \rho_r$ is the energy density for
both non relativistic (dust) and relativistic (radiation) matter
respectively. Regarding the dynamical system corresponding to the
total cosmological system, in addition to the variables introduced
in Eq. (\ref{xvariables}), the following two variables must be
added,
\begin{align}
\centering
\label{xrho}
x_5&=\frac{\kappa^2\rho_r}{3FH^2}&x_6&=\frac{\kappa^2\rho_m}{3FH^2}\, ,
\end{align}
so by taking these two into account, the Friedmann constraint that
must be satisfied by all the trajectories in the phase space,
becomes,
\begin{equation}
\centering \label{motion5} x_1+x_2+x_3+x_4+x_5+x_6 = 1\, .
\end{equation}
Thus, the new dynamical system corresponding to the cosmological
system at hand is now written as,
\begin{equation}
\centering
\label{Dx1}
\frac{dx_1}{dN}=x_1(x_1+3-x_3)+2(x_3-2)(1-2x_4)+4x_5+3x_6\, ,
\end{equation}
\begin{equation}
\centering
\label{Dx2}
\frac{dx_2}{dN}=x_2(x_1-2x_3+4)-4(x_3-2)+m\, ,
\end{equation}
\begin{equation}
\centering
\label{Dx3}
\frac{dx_3}{dN}=-m-2(x_3-2)^2\, ,
\end{equation}
\begin{equation}
\centering
\label{Dx4}
\frac{dx_4}{dN}=x_4(x_1+2x_3-4)\, ,
\end{equation}
\begin{equation}
\centering
\label{Dx5}
\frac{dx_5}{dN}=x_5(x_1-2x_3)\, ,
\end{equation}
\begin{equation}
\centering \label{Dx6} \frac{dx_6}{dN}=x_6(x_1-2x_3+1)\, .
\end{equation}
As mentioned previously, the system is autonomous only if $m$ is
specified accordingly. As in the previous section, we shall choose
the values $m = 0$ for the (quasi) de Sitter expansion, $m =
-\dfrac{9}{2}$ for the matter-dominated era, and $m = -8$ for the
radiation-dominated era. In the case at hand, the radiation and
matter domination eras can be realized by the corresponding
perfect fluids, and not necessarily from the $f(R)$ gravity.

The equilibrium points of the dynamical system are now increased
to ten, and these are,
\begin{align}
\centering \label{equilibrium2} &\phi_*^1 =\left( 3-\sqrt{-2m} ,
\dfrac{1}{3} \left(-m -2\sqrt{-2m} \right) , 2
-\frac{\sqrt{-2m}}{2} , 0 , 0 , \dfrac{1}{6} \left( 2m -24
+13\sqrt{-2m} \right) \right) \\ \notag &\phi_*^2= \left(
4-\sqrt{-2m} , -\dfrac{m + 2\sqrt{-2m}}{4} ,
2-\frac{\sqrt{-2m}}{2} , 0 , \dfrac{1}{4} \left( m -20
+8\sqrt{-2m} \right) , 0 \right) \\ \notag &\phi_*^3 =\left( 3
+\sqrt{-2m} , \dfrac{1}{3} \left(m -2\sqrt{-2m} \right) , 2
+\frac{\sqrt{-2m}}{2} , 0 , 0 , \dfrac{1}{6} \left( 2m -24
-13\sqrt{-2m} \right) \right) \\ \notag &\phi_*^4= \left(
4+\sqrt{-2m} , -\dfrac{m - 2\sqrt{-2m}}{4} ,
2+\frac{\sqrt{-2m}}{2} , 0 , \dfrac{1}{4} \left( m -20
-8\sqrt{-2m} \right) , 0 \right) \\ \notag &\phi_*^5= \left(
\frac{1}{4} \left( -2 -\sqrt{-2m} -\sqrt{4 -2m +20 \sqrt{-2m}}
\right) , \frac{1}{4} \left(-2 +3\sqrt{-2m} +\sqrt{4 -2m
+20\sqrt{-2m}} \right) , 2-\frac{\sqrt{-2m}}{2} , 0 , 0 , 0
\right) \\ \notag &\phi_*^6 =\left( \frac{1}{4} \left( -2
-\sqrt{-2m} +\sqrt{4 -2m +20 \sqrt{-2m}} \right) , \frac{1}{4}
\left(-2 +3\sqrt{-2m} -\sqrt{4 -2m +20\sqrt{-2m}} \right) ,
2-\frac{\sqrt{-2m}}{2} , 0 , 0 , 0 \right) \\ \notag &\phi_*^7=
\left( \frac{1}{4} \left( -2 +\sqrt{-2m} -\sqrt{4 -2m +20
\sqrt{-2m}} \right) , \frac{1}{4} \left(-2 -3\sqrt{-2m} +\sqrt{4
-2m +20\sqrt{-2m}} \right) , 2 +\frac{\sqrt{-2m}}{2} , 0 , 0 , 0
\right) \\ \notag &\phi_*^8= \left( \frac{1}{4} \left( -2
+\sqrt{-2m} +\sqrt{4 -2m +20 \sqrt{-2m}} \right) , \frac{1}{4}
\left(-2 -3\sqrt{-2m} -\sqrt{4 -2m +20\sqrt{-2m}} \right) , 2
+\frac{\sqrt{-2m}}{2} , 0 , 0 , 0 \right) \\ \notag &\phi_*^9=
\left( \sqrt{-2m} , \frac{1}{4} \left(-4 +\sqrt{-2m} \right) ,
2-\frac{\sqrt{-2m}}{2} , -\frac{3 \sqrt{-2m}}{4} , 0 , 0 \right)
\\ \notag &\phi_*^{10}= \left( -\sqrt{-2m} , -\dfrac{1}{4} \left( 4
+\sqrt{-2m} \right) , 2 +\dfrac{\sqrt{-2m}}{2} , \dfrac{3
\sqrt{-2m}}{4} , 0 , 0 \right)\, .
\end{align}
In the presence of the matter fields, four extra fixed points
emerge for general values of $m$, however -as was the case with
the vacuum $f(R)$ model- many points may coincide while other may
be discarded as being intrinsically unrealistic. In fact, at least
two of them -namely $\phi_*^7$ and $\phi_*^8$- will indeed be
discarded since these are complex. All the remaining eight are
viable, as they fulfill the Friedmann constraint of Eq.
(\ref{motion5}).

The Jacobian for the dynamical system of Eqs.
(\ref{Dx1})-(\ref{Dx5}) reads,
\begin{equation}
\centering
\label{Jacobian2}
J=
 \left(
\begin{array}{cccccc}
 2 x_1^* -x_3^* +3 & 0 & 2 ( 1 -2x_4^* ) -x_1^* & -4 ( x_3^* -2 ) & 4 & 3 \\
 x_2^* & x_1^* -2 x_3^* +4 & -2 x_2^* -4 & 0 & 0 & 0 \\
 0 & 0 & -4 ( x_3^* -2 ) & 0 & 0 & 0 \\
 x_4^* & 0 & 2 x_4^* & x_1^* +2 x_3^* -4 & 0 & 0 \\
 x_5^* & 0 & -2 x_5^* & 0 & x_1^* -2 x_3^* & 0 \\
 x_6^* & 0 & -2 x_6^* & 0 & 0 & x_1^* -2 x_3^* +1 \\
\end{array}
\right) \, ,
\end{equation}
again seeming independent of $m$, however the dependence of the
fixed points on the parameter $m$ brings up the indirect
dependence of the Jacobian as well, thus the position and the
stability properties of the equilibria alter upon changes of the
free parameter $m$. Dealing with the general case of linearizing
around all ten fixed points (for $m \leq 0$) is complicated, yet
the results are presented in table \ref{table2} -again, we have
omitted two points that are complex, hence of no physical meaning.
We observe that only one of the equilibria is asymptotically
stable for all values of $m$, hence we naturally expect the
solutions to be drawn to it. Not surprisingly, it is a reflection
of the equilibrium for the vacuum case, where $x_4$ takes a
nonzero value, while $x_5$ and $x_6$ are zero, hence the energy
densities of dust and radiation are bound to decrease.
\vspace{5mm}
\begin{table}[h!]
  \begin{center}
    \caption{\emph{\textbf{Equilibrium Points for the $f(R)$-perfect fluids autonomous system}}}
    \label{table2}
    \begin{tabular}{ |p{2cm}||p{6.9cm}|p{3cm}|p{4cm}|  }
     \hline
      \textbf{ Fixed Point} & \textbf{Eigenvalues} & \textbf{Characterization} & \textbf{Centre Manifolds}  \\
           \hline
 Fixed Point   & Eigenvalues & Characterization & Centre Manifolds \\
 \hline
 $\phi_*^1$ & $3$ , $-1,\dfrac{1}{4} \left(14 -3 \sqrt{-2m}-\sqrt{4 -2m +20 \sqrt{-2m}}\right)$ , $\dfrac{1}{4} \left(14 -3\sqrt{-2m} +\sqrt{4 -2m +20 \sqrt{-2m}}\right)$ , $3 -2 \sqrt{-2m}$ and $2 \sqrt{-2m}$ & Saddle & For $m=0$ and $m=-\dfrac{9}{8}$ and $m = - \dfrac{5}{2}$. \\
 $\phi_*^2$ & $4$ , $1$ , $\dfrac{1}{4} \left(18 -3\sqrt{-2m} -\sqrt{4 -2m +20\sqrt{-2m}}\right)$ , $\dfrac{1}{4} \left(18 -3 \sqrt{-2m} +\sqrt{4 -2m +20 \sqrt{-2m}}\right)$ , $4-2 \sqrt{-2m}$ and $2 \sqrt{2} \sqrt{-m}$ & Saddle & For $m=0$, $m = -2$ and $m = -4.80816$. \\
 $\phi_*^3$ & $3$ , $-1$ , $3 +2 \sqrt{-2m}$ , $\dfrac{1}{4} \left(14 +3\sqrt{-2m} -\sqrt{-2m -20\sqrt{-2m}+4}\right)$ , $\dfrac{1}{4} \left(14 +3\sqrt{-2m} +\sqrt{4 -2m -20\sqrt{-2m}}\right)$ and $-2\sqrt{-2m}$ & Generalized Saddle & For all $m=0$. \\
 $\phi_*^4$ & $4$ , $1$ , $4 +2\sqrt{-2m}$ , $\dfrac{1}{4} \left(18 +3\sqrt{-2m} -\sqrt{4 -2m -20\sqrt{-2m}}\right)$ , $\dfrac{1}{4} \left(18 +3\sqrt{-2m} +\sqrt{4 -2m -20\sqrt{-2m}}\right)$ and $-2\sqrt{-2m}$ & Generalized Saddle & For all $m=0$. \\
 $\phi_*^5$ & $\dfrac{1}{4} \left(-2 -5\sqrt{-2m}-\sqrt{4 -2m +20 \sqrt{-2m}}\right)$ , $\dfrac{1}{4} \left(-18 +3\sqrt{-2m} -\sqrt{4 -2m +20 \sqrt{-2m}}\right)$ , $\dfrac{1}{4} \left(-14 +3 \sqrt{-2m} -\sqrt{4 -2m +20\sqrt{-2m}}\right)$ , $\dfrac{1}{4} \left(-2 +3 \sqrt{-2m} -\sqrt{4 -2m +20 \sqrt{-2m}}\right)$ , $-\sqrt{1 + 5\sqrt{-2m} -\dfrac{m}{2}}$  and $2\sqrt{-2m}$ & Saddle & For $m=0$ and $m = -8$. \\
 $\phi_*^6$ & $\dfrac{1}{4} \left(-2 -5\sqrt{-2m} +\sqrt{4 -2m +20\sqrt{-2m}}\right)$ , $\dfrac{1}{4} \left(-18 +3\sqrt{-2m} +\sqrt{4 -2m +20\sqrt{-2m}} \right)$ , $\dfrac{1}{4} \left(-14 +3\sqrt{-2m} +\sqrt{4 -2m +20\sqrt{-2m}}\right)$ , $\dfrac{1}{4} \left(-2 +3\sqrt{-2m} +\sqrt{4 -2m +20\sqrt{-2m}}\right)$ , $\sqrt{1 +5\sqrt{-2m} -\frac{m}{2}}$ and $2 \sqrt{-2m}$ & Saddle & For $m=0$, $m = -2$ and $m = -4.80816$. \\
 $\phi_*^9$ & $-4 +2\sqrt{-2m}$ , $-3 +2\sqrt{-2m}$ , $2\sqrt{-2m}$ , $s_1$ , $s_2$ and $s_3$\rlap{*} & Saddle turning to Unstable Node & For $m=0$, $m = -\dfrac{9}{8}$ and $m = -2$. \\
 $\phi_*^{10}$ & $-4 -2\sqrt{-2m}$ , $-3 -2\sqrt{-2m}$ , $-2\sqrt{-2m}$ , $\tilde{s}_1$ , $\tilde{s}_2$ and $\tilde{s}_3$\rlap{**} & Stable Node-Focus & For $m=0$, $m = -6.52827$. \\
 \hline
 \multicolumn{4}{@{}l}{* Roots of polynomial $s^3 + \left(-18\sqrt{-2m} -4\right) s^2 +\left(32\sqrt{-2m} -208m\right) s +384m \sqrt{-2m}$.} \\
 \multicolumn{4}{@{}l}{**  Roots of polynomial $\tilde{s}^3 +\left(18\sqrt{-2m} -4\right) \tilde{s}^2 +\left(-32 \sqrt{-2m} -208m\right) \tilde{s} -384m \sqrt{-2m}$.}
\end{tabular}
  \end{center}
\end{table}
What we should notice, before moving towards the specific
cosmological eras, specified by the values of $m$, is that
whatever said for Eq. (\ref{dx3}) holds for Eq. (\ref{Dx3}), as
they are identical; these equations are decoupled from the system
and integrated analytically producing the solutions noted in Fig.
\ref{fig:x3_integrable}. This is particularly important for this
case, as it may also decouple Eqs. (\ref{Dx5}) and (\ref{Dx6}). As
the rates of change of $x_5$ and $x_6$ dependent only on $x_1$ and
$x_3$, they will converge or diverge depending on whether the
expression $x_1-2x_3 +1$ is smaller than $-1$. Given that $x_{3}$
converges rapidly at $2 +\dfrac{\sqrt{-2m}}{2}$ if $x_{3} > 2
-\dfrac{\sqrt{-2m}}{2}$, then we easily see that $x_1$ is also
bounded by,
\begin{equation*}
x_1 < 2 + \sqrt{-2m} \, .
\end{equation*}
Interestingly, the value $x_1$ takes at the stable fixed point
$\phi_*^{10}$ is $-2 \left( 2+\sqrt{-2m} \right)$, which is below
that bound.

Ensuring that our initial conditions respect $x_1 < 2 +
\sqrt{-2m}$, $x_{3} > 2 -\dfrac{\sqrt{-2m}}{2}$ and $x_4$, $x_5$,
$x_6 \geq 0$, as well as the Friedmann constraint, we ensure that
$\phi_*^{10}$ is reached normally. We clearly see from that how
the centre and unstable manifolds divide the phase space -yet
again- and force specific initial conditions in order for a viable
solution to exist.

\subsection{The case of de Sitter Expansion: $m=0$}

Now let us study the subspace of the total phase space of the
cosmological system that corresponds to (quasi) de Sitter
solutions, so for $m=0$. In this case, the fixed points of the
dynamical system are,
\begin{align}
\centering \phi_*^1& (3,0,2,0,0,-4)&\phi_*^2&
(4,0,2,0,-5,0)&\phi_*^5& (-1,0,2,0,0,0)&\phi_*^6 (0,-1,2,0,0,0)\,
.
\end{align}
As it was mentioned earlier, the total number of equilibrium
points is decreased and essentially there exist only four, out of
total eight, and the rest are identical to these. Concerning the
eigenvalues of Eq. (\ref{Jacobian2}) for each of these points, we
can mention the following: Equilibrium $\phi_*^1$ yields $\{4, 3,
3, 3, -1, 0\}$, being a saddle with $\vec{v}_{-1,6} = (-3, 0, 0,
0, 0, 4)$, $\vec{v}_{2} =(0, 1, 0, 0, 0, 0)$ and $\vec{v}_{4} =(0,
0, 0, 1, 0, 0)$ being tangent to the degenerate unstable
manifolds, $\vec{v}_{1,-5,6} =(1, 0, 0, 0, -5, 4)$ to the stable
manifold, and $\vec{v}_{0} =(-6, -4, -3, 0, 0, 13)$ to a centre
manifold. Equilibrium $\phi_*^2$ is characterized as a degenerate
unstable node, with its eigenvalues being $\{5, 4, 4, 4, 1, 0\}$
and $\vec{v}_{0} = (3,-4,1,0,0,0)$ being tangent to its only
centre manifold. Equilibrium $\phi_*^5$ is a degenerate stable
node, as the eigenvalues of Eq. (\ref{Jacobian2}) for it are
$\{-5, -4, -1, -1, -1, 0\}$, with $\vec{v}_{1}=(1,0,0,0,0,0)$,
$\vec{v}_{2}=(0,1,0,0,0,0)$ and $\vec{v}_{4}=(0,0,0,1,0,0)$
denoting the degenerate stable manifolds and $\vec{v}_{0} =(-6,
-4, -3, 0, 0, 13)$ the corresponding centre manifold. Finally,
$\phi_*^6$ is a saddle dominated by centre manifolds, since the
eigenvalues are $\{-4, -3, 1, 0, 0, 0\}$. Notably, three centre
manifolds appear tangent to directions
$\vec{v}_{-1,-3}=(-2,0,-1,0,0,0)$, $\vec{v}_2=(0,1,0,0,0,0)$ and
$\vec{v}_4=(0,0,0,1,0,0)$.

Consequently, the solutions are expected to converge to $\phi_*^5$
as long as the discussed bounds are considered. As in the previous
case, we can see that $x_{3} = 2$ even for very small values of
the dynamical evolving variable $N$. In this case, the system of
Eqs. (\ref{Dx1}), (\ref{Dx2}), (\ref{Dx4}) (\ref{Dx5}) and
(\ref{Dx6}) is simplified and can be integrated, yielding,
\begin{align}
x_1 (N) =& \dfrac{ x_{1}(0) e^N }{1 + x_{1}(0) \left( 1-e^N \right)} \, , \;\;\;
x_2 (N) = \dfrac{ x_{2}(0) }{1 + x_{1}(0) \left( 1-e^N \right)} \, , \;\;\;
x_4 (N) = \dfrac{ x_{4}(0) }{1 + x_{1}(0) \left( 1-e^N \right)} \, , \\ \notag
&x_5 (N) = \dfrac{ x_{5}(0) e^{-4N} }{1 + x_{1}(0) \left( 1-e^N \right)} \;\;\; \text{and} \;\;\;
x_6 (N) = \dfrac{ x_{6}(0) e^{-3N} }{1 + x_{1}(0) \left( 1-e^N \right)} \, ,
\end{align}
where $x_{1}(0)$, $x_{2}(0)$, $x_{4}(0)$ $x_{5}(0)$ and $x_{6}(0)$
are the initial values. All solutions are naturally led to the
fixed point. On the $x_1 - x_2 -x_4$ hyperplane, the parametric
solution of Eqs. (\ref{integrals}) is valid, defining the
respective stable manifold -as in the vacuum case. The overall
behavior can easily be seen in Fig. \ref{fig:first_deSitter},
where the solutions do indeed follow this pattern, attracted and
led to equilibrium point $\phi_5^*$. Obviously, any solutions
whose initial conditions do not respect the afore-mentioned
limitations are repelled away, as in the vacuum case.

\subsection{The case of Matter-Dominated Cosmologies: $m=-\frac{9}{2}$}

Setting $m = -\dfrac{9}{2}$, we transcend to the matter-dominated
era, where all eight equilibria appear, which are,
\begin{align}
\centering &\phi_*^1 =\left( 0,-\frac{1}{2},\frac{1}{2},0,0,1
\right)\\ \notag &\phi_*^2 =\left(
1,-\frac{3}{8},\frac{1}{2},0,-\frac{1}{8},0 \right)\\ \notag
&\phi_*^3 =\left( 6,\frac{7}{2},\frac{7}{2},0,0,-12 \right)\\
\notag &\phi_*^4 =\left(
7,\frac{21}{8},\frac{7}{2},0,-\frac{97}{8},0 \right)\\ \notag
&\phi_*^5 =\left(
\frac{1}{4}\left(5-\sqrt{73}\right),\frac{1}{4}\left(7+\sqrt{73}
\right),\frac{1}{2},0,0,0 \right)\\ \notag &\phi_*^6 =\left(
\frac{1}{4}\left(-5+\sqrt{73}\right),\frac{1}{4}\left(7-\sqrt{73}
\right),\frac{1}{2},0,0,0 \right)\\ \notag &\phi_*^7 =\left(
3,-\frac{1}{4},\frac{1}{2},-\frac{9}{4},0,0 \right)\\ \notag
&\phi_*^8 =\left( -3,-\frac{7}{4},\frac{7}{2}, \frac{9}{4},0,0
\right)\, ,
\end{align}
Concerning their stability, we can draw the following conclusions
from the linearization of the system in the neighborhood of each.
The fixed points $\phi_*^1$ and $\phi_*^2$ exhibit eigenvalues $\{
6, 3.386, 3, -3, -1, -0.886001 \}$ and $\{ 6, 4.386, 4, -2, 1,
0.113999 \}$, hence they are typical saddles, whereas for
$\phi_*^3$ and $\phi_*^4$ we obtain eigenvalues $\{ 9, -6, 5.75
+1.71391 i, 5.75 -1.71391 i, 3, -1 \}$ and $\{ 10, 6.75 +1.71391
i, 6.75 -1.71391 i, -6, 4, 1 \}$ respectively, so these are
characterized as generalized saddles. The linearizations around
$\phi_*^5$ result to the eigenvalues $\{ -6.386, 6, -4.386,
-4.272, -3.386, -0.386001 \}$, while for $\phi_*^6$ we have $\{ 6,
4.272, 3.886, -2.114, 0.886001, -0.113999 \}$, thus they are both
saddles. Finally, the Jacobian in the neighborhood of the
equilibrium point $\phi_*^7$ yields eigenvalues $\{ 6.386, 6, 6 ,3
, 2.114, 2 \}$, making it an unstable node, whereas in the
neighborhood of equilibrium $\phi_*^8$ we get, $\{ -10, -9, -6,
-6, -3.25 + 1.71391i, -3.25 - 1.71391i \}$, making $\phi_*^8$ a
stable node-focus.  Both possess a degeneracy over the $\vec{v}_2
= ( 0,1,0,0,0,0 )$ direction.

As in the vacuum case, the solutions whose initial values follow
the restrictions are attracted to the plane defined orthogonally
to $\vec{v}_{-1,2,4} = (-1.44444, 0.444444, 0, 1, 0, 0 )$ and
there they oscillate towards the stable fixed point. The
oscillations depend of the real and the imaginary part of the
sixth and fifth eigenvalue of $\phi_*^8$. Subsequently, the
cyclical behavior is very small in comparison to the attraction.
Thus the $x_1 - x_2 - x_4$ slice of the phase space of the system
is identical to that of the previous system, as it is shown in
Fig. \ref{fig:first_Matter}.

The nonzero value of $x_4$ is preserved at this fixed point,
regardless of the perfect fluids present, whose cosmological
parameters are drawn to zero, usually within a few $e$-foldings
-as the numerical solutions show. As a result, it is the thermal
term $\sim \alpha H^4$ that characterized the respective viable
cosmological solutions, rather than the matter fields themselves.
However, the values of $x_1$ and $x_2$ differ significantly from
the corresponding viable solutions in the vacuum case, indicating
the role the matter fields play in the functional form of the
$f(R)$ gravity actually.

\subsection{The case of Radiation-Dominated Cosmologies: $m=-8$}

In much the same manner, when we set $m = -8$ that corresponds to
the radiation-dominated era, we have the following eight
equilibrium points,
\begin{align}
\centering &\phi_*^1 =\left( -1, 0, 0, 0, 0, 2 \right)\\ \notag &
\phi_*^2 =\left( 0, 0, 0, 0, 1, 0 \right)\\ \notag & \phi_*^3=
\left( 7, \dfrac{16}{3}, 4, 0, 0, -\dfrac{46}{3} \right)\\ \notag
& \phi_*^4 =\left( 8, 4, 4, 0, -15, 0 \right)\\ \notag & \phi_*^5=
\left( -4, 5, 0, 0, 0, 0 \right)\\ \notag & \phi_*^6 =\left( 1, 0,
0, 0, 0, 0 \right)\\ \notag & \phi_*^7 =\left( 4, 0, 0, -3, 0, 0
\right)\\ \notag & \phi_*^8= \left( -4, -2, 4, 3, 0, 0 \right)\, ,
\end{align}
Given the Jacobian of Eq. (\ref{Jacobian2}) for each of them, we
can demonstrate their stability characterization. The fixed points
$\phi_*^1$ and $\phi_2^*$ are saddles, as the respective
eigenvalues are $\{8, -5, 3, 3, -2, -1\}$ and $\{8, 4, 4, -4, 1,
-1\}$. It is noteworthy that direction $\vec{v}_2 = (0,1,0,0,0,0)$
denotes a degenerate unstable manifold for both the fixed points.
The fixed points $\phi_3^*$ and $\phi_*^4$ are characterized as
generalized saddles, since they exhibit eigenvalues $\{ 11, -8,
6.5 +1.93649 i, 6.5 -1.93649 i, 3, -1 \}$ and $\{ 12, -8, 7.5
+1.93649 i, 7.5 -1.93649 i, 4, 1 \}$ respectively, while
$\phi_*^6$ is also a typical saddle with eigenvalues $\{ 8, 5, 5,
-3, 2, 1 \}$ with a degeneracy over $\vec{v}_1 = (1,0,0,0,0,0)$
and $\vec{v}_2 = (0,1,0,0,0,0)$ directions.  The fixed point
$\phi_*^5$ is yet another saddle, which however possesses a centre
manifold, as its eigenvalues are $\{ -8, 8, -5, -4, -3, 0 \}$. The
centre manifold is tangent to $\vec{v}_2 = (0,1,0,0,0,0)$.
Equilibrium $\phi_*^7$ is characterized by eigenvalues $\{ 8, 8,
8, 5, 4, 3 \}$ and hence it is thus a degenerate unstable node.
Finally, equilibrium $\phi_*^8$ is a stable node-focus, as the
linearized system around it yields $\{ -12, -11, -8, -8,
-4.5+1.93649 i, -4.5-1.93649 i \}$. Interestingly, directions
$\vec{v}_{-3,4} =(0,0,-4,3,0,0)$ and $\vec{v}_2 =(0,1,0,0,0,0)$
are tangent to degenerate unstable manifolds, due to the equality
of the respective eigenvalues.

Thus in general, the numerical behavior of the trajectories in the
case at hand, are similar to those depicted in Fig.
\ref{fig:first_Radiation}.

\subsection{Non-autonomous Dynamical System in the Presence of
Perfect Fluids}

Finally, it is worth extending the Big Rip discussion of a
previous section, in the case that perfect matter fluids are
present, in which case the dynamical system becomes,
\begin{align}\label{finitesystem2}
\frac{\mathrm{d}x_1}{\mathrm{dN}} &= x_1(x_1+3-x_3) +2(x_3-2)(1-2x_4) + 4 x_5 + 3 x_6 \, , \\ \notag
\frac{\mathrm{d}x_2}{\mathrm{dN}} &= x_2(x_1-2x_3+4) -4(x_3-2) -\dfrac{\beta ( \beta+1 )}{ (\beta - 1)^2 \left( N_c - N \right)^2 } \, , \\ \notag
\frac{\mathrm{d}x_3}{\mathrm{dN}} &= \dfrac{\beta ( \beta+1 )}{ (\beta - 1)^2 \left( N_c - N \right)^2 } -2(x_3-2)^2  \, , \\ \notag
\frac{\mathrm{d}x_4}{\mathrm{dN}} &= x_4(x_1+2x_3-4) \, , \\ \notag
\frac{dx_5}{dN} &= x_5(x_1-2x_3) \;\;\; \text{and} \\ \notag
\frac{dx_6}{dN} &= x_6(x_1-2x_3+1) \, ,
\end{align}
where $\beta$ can again take the values considered in Section III.

In general, the results recovered for the vacuum case are true in
this case as well. Considering the Big Rip singularity, the system
for any $N_c > 0$ it is attracted initially and repelled later on,
as $N \rightarrow N_c$; the repulsion occurs on a straight line
that we have traced analytically in Eqs. (\ref{bigripattractor}).
Therefore the perfect fluids do not affect the behavior of the
dynamical system near the Big Rip singularity, a behavior that was
anticipated.

The issue of finite-time singularities occurring in a cosmological
framework, by using the dynamical systems approach was also
addressed in Ref. \cite{Odintsov:2018uaw}. In the present analysis
we used a more direct approach to see whether a Big Rip
singularity may be realized by the thermal term corrected $f(R)$
gravity phase space. Our findings indicated that indeed a
late-time de Sitter fixed point exists, but there is a difference
with the non-thermal-corrected $f(R)$ gravity, the de Sitter fixed
point of the thermally corrected $f(R)$ theory is highly unstable,
and we pinpointed the source of the instability, which is the fact
that for all the initial conditions, the fixed point is unstable
near the starting point of the trajectory.

\section{Quantitative Analysis of the Thermal Term Effects on the Late-time Era}

Let us now follow another research line and we examine directly
the effects of the thermal terms on the late-time evolution of
$f(R)$ gravity, by investigating the behavior of several late-time
related observational quantities. We shall make a direct
comparison of the ordinary $f(R)$ gravity and the $f(R)$ gravity
with thermal correction terms, in the presence of matter and
radiation perfect fluids.
\begin{figure}[h!]
\centering
\includegraphics[width=20pc]{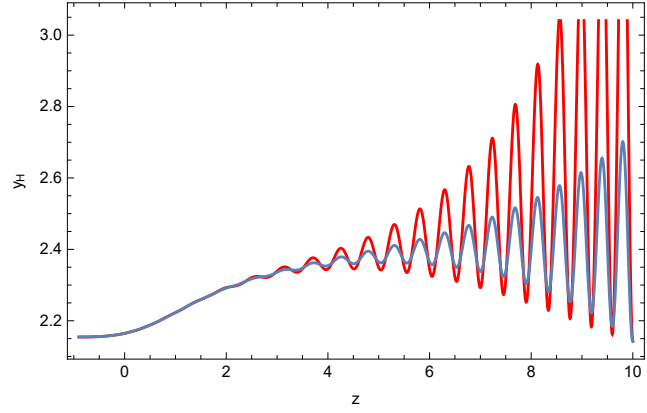}
\caption{Plot of the behavior of the statefinder $y_H(z)$ as a
function of the redshift for the ordinary $f(R)$ gravity (blue
curve) and for the thermally corrected $f(R)$ gravity (red curve),
for $\alpha=10^{98}$.} \label{plot1newsec}
\end{figure}
The corresponding gravitational action is,
\begin{equation}
\centering \label{action}
S=\int{d^4x\sqrt{-g}\left(\frac{f(R)}{2\kappa^2}+\mathcal{L}_{eff}+\mathcal{L}_m\right)}\,
,
\end{equation}
where the term $\mathcal{L}_m$ denotes the Lagrangian density of
the radiation and non-relativistic perfect matter fluids, and the
term $\mathcal{L}_{eff}$ denotes the Lagrangian density
responsible for the thermal effects. We shall assume that the
background metric is a flat FRW, and also we shall additionally
assume that the $f(R)$ gravity model is of the form
\cite{Odintsov:2020nwm},
\begin{equation}
\centering \label{F(R)}
f(R)=R+\frac{R^2}{M^2}-\gamma\Lambda\left(\frac{R}{3m_s^2}\right)^\delta\,
,
\end{equation}
where $M=1.5\cdot10^{-5}\frac{50}{N}M_p$, $N$ is the $e$-foldings
number, $\Lambda$ is the cosmological constant, multiplied by
parameter $\gamma$, $m_s$ is the mass scale defined as
$m_s^2=H_0^2\Omega_m^0$ and $\delta$ takes values from the
interval $0<\delta<1$. The reason for choosing such an $f(R)$
gravity is that it can generate a quite viable late-time era, as
it was shown in detail in Ref. \cite{Odintsov:2020nwm}, see also
\cite{Odintsov:2020qyw}.
\begin{figure}[h!]
\centering
\includegraphics[width=20pc]{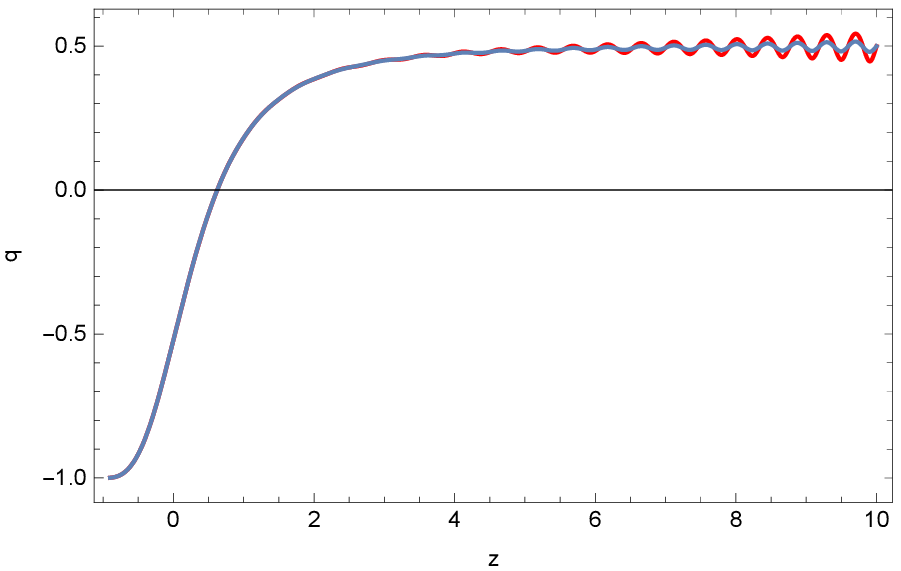}
\includegraphics[width=20pc]{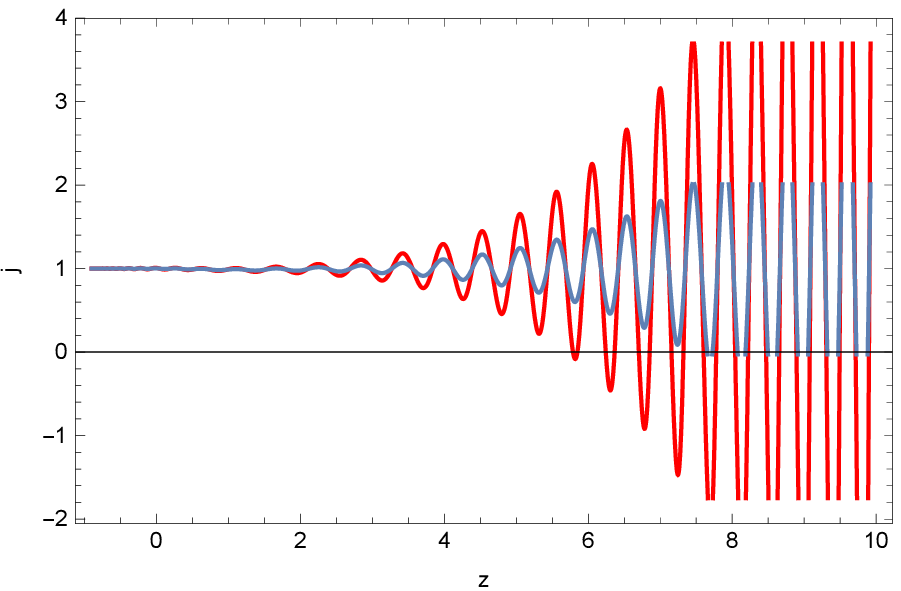}
\includegraphics[width=20pc]{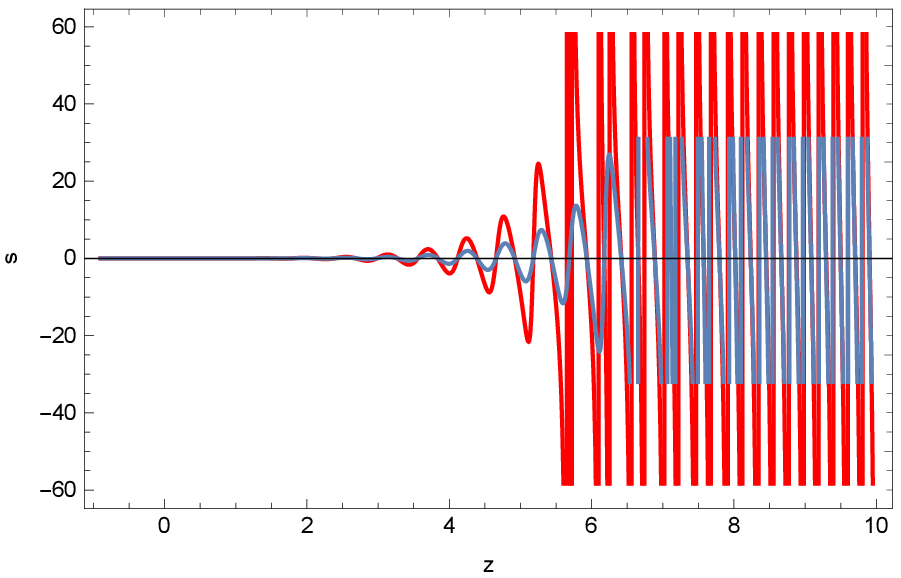}
\includegraphics[width=20pc]{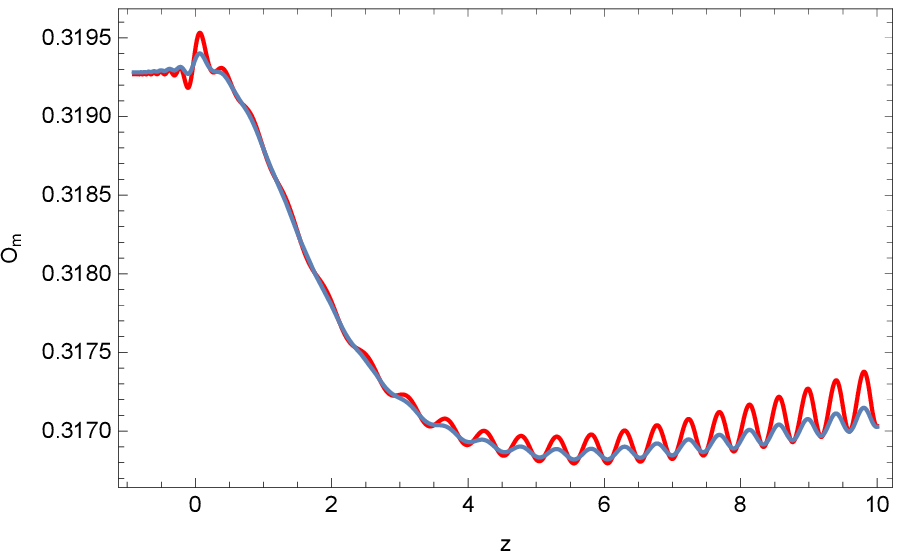}
\caption{Plot of the deceleration parameter (upper left), the jerk
(upper right), the snap (bottom left) and $Om(z)$ (bottom right)
as a function of the redshift for the ordinary $f(R)$ gravity
(blue curves) and for the thermally corrected $f(R)$ gravity (red
curves), for $\alpha=10^{98}$.} \label{plot2newsec}
\end{figure}
As it was shown in the aforementioned references, the late-time
era is mainly controlled by the term $R^\delta$, which becomes
dominant during the dark energy era, for $0<\delta<1$. The
Friedmann equation in the presence of the thermal term $\sim
\alpha H^4$ is,
\begin{equation}
\centering \label{motion1newsec}
\frac{3FH^2}{\kappa^2}=\rho_{m}+\alpha
H^4+\frac{RF-f}{2\kappa^2}-\frac{3H\dot F}{\kappa^2}\, ,
\end{equation}
where $\rho_{m}$ denotes the energy density of the radiation and
non-relativistic matter perfect fluids. We shall express the
Friedmann equation in terms of the redshift $z$, which, by
assuming that the present time scale factor is equal to unity, is
defined as follows,
\begin{equation}
\centering \label{z} 1+z=\frac{1}{a(t)}\, ,
\end{equation}
and also we shall make use of the following differentiation rule,
\begin{equation}
\centering \label{dz} \frac{d}{dt}=-H(z)(1+z)\frac{d}{dz}\, .
\end{equation}
Hence, the Friedmann equation is written in terms of the redshift
as follows,
\begin{equation}
\centering \label{motion3newsec}
\frac{3FH^2}{\kappa^2}=\rho_m+\frac{RF-f}{2\kappa^2}+\frac{3H^2(1+z)R'F_{R}}{\kappa^2}+\alpha
H^4\, ,
\end{equation}
where we took into account the fact that $\dot F=F_{R}\dot R$
and $\dot R=-H(1+z)R'$, with the ``prime'' denoting
differentiation with respect to the redshift for simplicity
hereafter. As it can easily be inferred, the gravitational field
equations deviate from Einstein's description, hence the reason
why Friedman's equation has a different form. However, we can
perform a simple replacement which restores the original, or usual
form of Friedman's equations while simultaneously giving a
physical meaning to the so called dark energy fluid. Suppose that
all the geometric contributions in the Friedman equation
(\ref{motion3newsec}) is the density of dark energy, meaning that

\begin{equation}
\centering \label{rhoG} \rho_G=\frac{1}{2}\left(\frac{R}{\kappa
M}\right)^2+\frac{1-\delta}{2}\frac{\gamma\Lambda}{\kappa^2}\left(\frac{R}{3m_s^2}\right)^\delta+\alpha
H^4-\frac{3H\dot
R}{\kappa^2}\left(\frac{2}{M^2}+\frac{\gamma\Lambda\delta(1-\delta)}{(3m_s^2)^\delta}R^{\delta-2}\right)-\frac{3H^2}{\kappa^2}\left(\frac{2R}{M^2}-\frac{\gamma\Lambda\delta}{(3m_s^2)^\delta}R^{\delta-1}\right)\,
,
\end{equation}
where $G$ stands for the geometric terms and in order not to make
the equation lengthy, we used $\dot R$ but essentially is written
as,
\begin{equation}
\centering \label{dotR} \dot
R=-H(1+z)R'=6H(1+z)\left((1+z)(H'^2+HH'')-3HH'\right)\, .
\end{equation}
The same could be said about the pressure of the dark energy but
essentially, by treating it as a perfect fluid, the equation of
state (EoS) can be extracted from the continuity equation,
\begin{equation}
\centering \label{conteq} \dot\rho_G+3H\rho_G(1+\omega_G)=0\, ,
\end{equation}
but we shall introduce it in the following. As a result, Eq.
(\ref{motion3}) is now written as,
\begin{equation}
\centering \label{motion4newsec}
H^2=\frac{\kappa^2}{3}\left(\rho_m+\rho_G\right)\, ,
\end{equation}
This enables us to make use of simple statefinder functions in
order to study the dynamics of the model. Indeed, let us introduce
function $y_H(z)=\frac{\rho_G}{\rho_m^{(0)}}$ where $\rho_m^{(0)}$
is used in order to make this parameter dimensionless, and
$\rho_m^{(0)}$ is the dark matter energy density at present time.
Consequently, by using the usual Friedman equation, we have,
\begin{equation}
\centering \label{yH}
y_H(z)=\frac{H^2}{m_s^2}-(1+z)^3-\chi(1+z)^4\, ,
\end{equation}
where $\chi$ is the current ratio of the relativistic over
non-relativistic density. In the following we can replace Hubble's
parameter with the statefinder parameter $y_H$ and afterwards
solve Eq. (\ref{motion3newsec}) numerically by appropriately
designating the initial conditions. Furthermore, we shall compare
the ordinary $f(R)$ with the thermal effects corrected $f(R)$
gravity, for redshifts in the range $z=[0,10]$. Concerning dark
energy we introduce two statefinder parameters, the EoS parameter
$\omega_G$ mentioned before and the density parameter $\Omega_G$,
which are defined as follows,
\begin{align}
\centering \label{DE}
\omega_G&=-1+\frac{1+z}{3}\frac{d\ln(y_H)}{dz}&\Omega_G&=\frac{y_H}{y_H+(1+z)^3+\chi(1+z)^4}\,
,
\end{align}
and for the overall evolution, we have the following statefinder
parameters,

\begin{align}
\centering \label{statefinders} q&=-1-\frac{\dot
H}{H^2}&j&=\frac{\ddot
H}{H^3}-3q-2&s&=\frac{j-1}{3(q-\frac{1}{2})}&O_m(z)&=\frac{\left(\frac{H(z)}{H_0}\right)^2-1}{(1+z)^3-1}\,
,
\end{align}
which are the deceleration, jerk, snap parameter respectively,
while $O_m(z)$ is an auxiliary statefinder which its current value
is indicative of the matter density $\Omega_m$. Since every object
is written in terms of redshift, as the name statefinder suggests,
we mention for simplicity that the deceleration and jerk
parameters are written as function of redshift as shown below,
\begin{equation}
\centering \label{q} q=-1+\frac{(1+z)H'}{H} \, ,
\end{equation}

\begin{equation}
\centering \label{j}
j=1+\left(\frac{(1+z)H'}{H}\right)^2+(1+z)^2\frac{H''}{H}-2(1+z)\frac{H'}{H}\,
,
\end{equation}
With regard to the values of the cosmological parameters, we shall
assume that the Hubble rate is \cite{Aghanim:2018eyx},
\begin{equation}\label{H0today}
H_0=67.4\pm 0.5 \frac{km}{sec\times Mpc}\, ,
\end{equation}
so $H_0=67.4km/sec/Mpc$ or equivalently $H_0=1.37187\times
10^{-33}$eV, therefore $h\simeq 0.67$. Hence, we shall take into
account only the CMB based value of the Hubble rate only,
disregarding the Cepheid based value. In addition, according to
the CMB extracted observational data $\Omega_c h^2$ is,
\begin{equation}\label{codrdarkmatter}
\Omega_c h^2=0.12\pm 0.001\, ,
\end{equation}
which we also used earlier for the definition of the parameter
$m_s^2$. Furthermore the parameter $\Lambda$ appearing in Eq.
(\ref{F(R)}), will be taken equal to $\Lambda\simeq 11.895\times
10^{-67}$eV$^2$ and in addition, the parameter $m_s^2$ expressed
in eV $m_s^2\simeq 1.87101\times 10^{-67}$eV$^2$, while $M$ is
$M\simeq 3.04375\times 10^{22}$eV for $N\sim 60$.
\begin{figure}[h!]
\centering
\includegraphics[width=20pc]{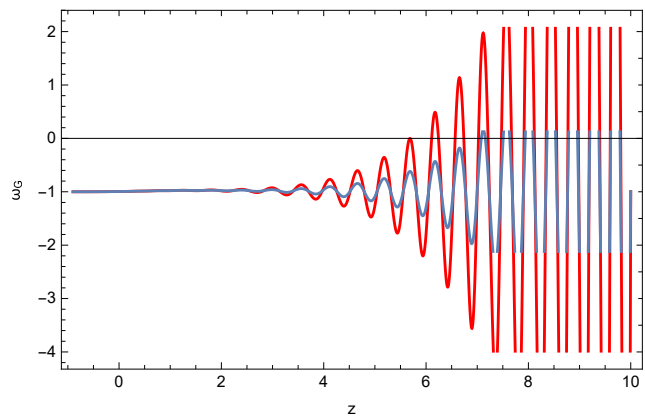}
\caption{Plot of the behavior of the dark energy EoS as a function
of the redshift for the ordinary $f(R)$ gravity (blue curve) and
for the thermally corrected $f(R)$ gravity (red curve), for
$\alpha=10^{98}$.} \label{plot3newsec}
\end{figure}
For the numerical study we shall use the following initial
conditions for the function $y_H(z)$
\cite{Odintsov:2020nwm,Bamba:2012qi},
\begin{align}\label{ft27}
& y_H(z=z_{f})=\frac{\Lambda}{3m_s^2}(1+\frac{1+z_{f}}{1000})\,
,\\ \notag &
\frac{dy_H}{dz}\Big{|}_{z=z_{f}}=\frac{\Lambda}{3m_s^2}\frac{1}{1000}\,
,
\end{align}
where $z_f$ denotes the final redshift $z_f=10$. The results of
our analysis indicated that the thermal terms effects are
significant only for extremely large values of the dimensionless
parameter $\alpha$, and specifically for $\alpha\geq 10^{98}$.
This can be seen in several statefinder parameters, for example in
Fig. \ref{plot1newsec} we plot the behavior of the statefinder
$y_H(z)$ as a function of the redshift for the ordinary $f(R)$
gravity (blue curve) and for the thermally corrected $f(R)$
gravity (red curve), for $\alpha=10^{98}$. It is apparent that the
dark energy oscillations are more pronounced for large redshifts,
for these large $\alpha$ values. The same results are obtained by
studying other statefinder quantities, for example in Fig.
\ref{plot2newsec} we present the plots of the deceleration
parameter (upper left), the jerk (upper right), the snap (bottom
left) and $Om(z)$ (bottom right) as a function of the redshift for
the ordinary $f(R)$ gravity (blue curves) and for the thermally
corrected $f(R)$ gravity (red curves), for $\alpha=10^{98}$.
Finally, differences between the ordinary $f(R)$ gravity (blue
curve) and for the thermally corrected $f(R)$ gravity (red curve)
can be found by studying the behavior of the dark energy EoS
parameter, as it can be seen in Fig. \ref{plot3newsec}. Finally,
the only cosmological parameter for which no differences occur
between the ordinary $f(R)$ gravity and the thermally corrected
$f(R)$ gravity, is the dark energy density parameter $\Omega_G$,
as it can be seen in Fig. \ref{plot4newsec}.
\begin{figure}[h!]
\centering
\includegraphics[width=20pc]{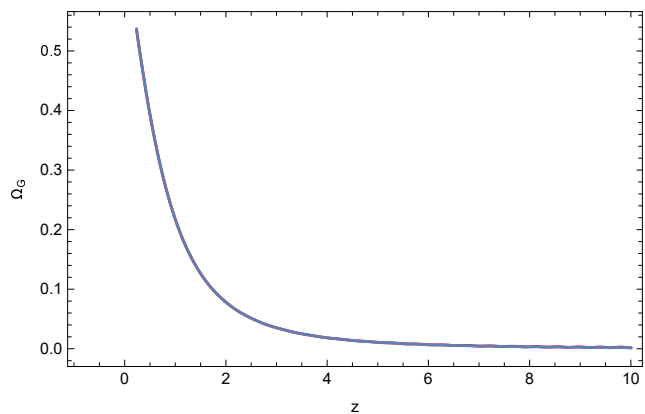}
\caption{Plot of the behavior of the dark energy density parameter
as a function of the redshift, for the ordinary $f(R)$ gravity
(blue curve) and for the thermally corrected $f(R)$ gravity (red
curve), for $\alpha=10^{98}$.} \label{plot4newsec}
\end{figure}
In conclusions, the results of our analysis indicate that the
ordinary $f(R)$ gravity and the thermally corrected $f(R)$ gravity
have differences only when the coefficient $\alpha$ of the thermal
corrections takes quite large values, and the differences can be
spotted in statefinders only for redshifts $z\geq 5$.

\section{Quantitative Discussion on the Results Obtained and Concluding Remarks}

In this paper we examined quantitatively the effects of a thermal
term $\sim \alpha H^4$ in the phase space of $f(R)$ gravity, in
the presence or not of perfect matter fluids. In order to perform
our quantitative analysis of the phase space, we have formulated
the theory as a dynamical system of four variables (vacuum case)
with respect to the $e$-foldings number. This system is similar to
the one proposed in Ref. \cite{Odintsov:2018uaw}, where the
authors dealt with the plain $f(R)$ theory. Our main difference is
that an additional dynamic variable exists representing the
effects of the thermal term $\sim \alpha H^4$. Our purpose, in
comparison to Ref. \cite{Odintsov:2018uaw}, was to unveil the
difference this term may bring up in the phase space behavior of
trajectories.

Choosing the Hubble rate such as to mimic the most important
cosmological eras, namely the de Sitter (or quasi-de Sitter)
accelerating expansion and the matter-dominated and radiation
dominated decelerating expansions, we ended up to an autonomous
dynamical system that was able to be studied qualitatively.
Basically, by focusing on these distinct cosmological eras, we
essentially studied the subspaces of the total phase space,
corresponding to these cosmological eras. Subsequently, we
proceeded by locating the equilibrium points for each of the three
cosmological eras, and thoroughly examined their stability
properties. Then, we numerically integrated the dynamical system
and we plotted the phase space trajectories to validate our
results. Notably, the (quasi-)de Sitter phase is dominated by
centre manifolds, however they can be isolated and the stability
of the equilibria can be addressed by means of the Hartman-Grobman
theorem. For the other two cosmological eras, the fixed points
were found to be hyperbolic, thus no such problem arose.

While in the (quasi-)de Sitter phase, the system reaches a stable
fixed point that is identical to the plain vacuum theory, meaning
that the extra thermal term does not play any crucial role during
this era. In that case, we reached viable cosmological solutions
with,
\begin{equation}
R = 12 H_0^2 \, ,
\end{equation}
in the de Sitter case, where $H(t) = H_0$, or
\begin{equation}
R = 12 H_0^2 -24 H_0 H_1 t +12 H_1^2 t^2 \, ,
\end{equation}
in the quasi-de Sitter case, where $H(t) = H_0 - H_1 t$, while,
\begin{equation}
 x_4 = 0 \, ,
\end{equation}
hence, the effect of the thermal fluid term is asymptotically zero
for all the stable trajectories.

However, when the Universe is described by either matter- or
radiation-dominated eras, the fixed points of the vacuum $f(R)$
theory become unstable and the extra thermal term has non-trivial
effects. In other words, the additional thermal term which was
added by hand in order to examine these eras, can cause dramatic
effects during the classical eras of the standard cosmological
model and the viable cosmological solutions are characterized by a
non-trivial thermal effects. In this case, for $m =
-\dfrac{9}{2}$, meaning that $H(t) = \dfrac{2}{3t}$, we have
\begin{equation}
R = \dfrac{28}{3t^2} \, ,
\end{equation}
and for $m = -8$ and thus $H(t) = \dfrac{1}{2t}$, we have
\begin{equation}
R = \dfrac{6}{t^2} \, ,
\end{equation}
so curvature decreases rapidly over time. As for the additional
thermal term, it increases rapidly over time as
\begin{equation}
\alpha H^4= \dfrac{27}{2\kappa^2} t^4 \, ,
\end{equation}
so it becomes more and more important as the viable
solution-trajectory in the phase space is attained. Thus clearly,
the thermal effects should not be present during the (quasi)
de-Sitter, matter and radiation domination eras, as we expected.
The effects of the thermal terms though, cause severe
instabilities to the dynamical system near a Big Rip singularity,
as we expected.

Introducing two typical perfect fluids, relativistic (radiation)
and non-relativistic (dust) matter as in the standard cosmological
model, the situation does not change. The two further dynamic
variables included are rapidly attracted to zero, hence the fixed
points of the vacuum theory are basically the same and the above
analysis holds true in this case too.

Choosing a different type of Hubble rate, of the type that models
the four types of finite-time singularities, we reach a
non-autonomous dynamical system, which cannot be analyzed
qualitatively in the same manner, mostly due to possessing no
invariant sets \textit{per se}. In this case, we attempted to
reduce the system and derive analytical solutions. Indeed, for the
case of the ``Big Rip'' singularity, an analytical solution is
obtained that matches the numerical solutions; these cannot
asymptotically reach the (quasi-)de Sitter fixed points as the
$e$-foldings number tends to infinity, since for any initial value
of $N_c$, the trajectories are repelled away as the singularity is
approached. Thus, as we already mentioned, the dynamical system of
$f(R)$ gravity has strong instabilities when it is studied near a
Big Rip singularity, as we actually expected.

Moreover, we examined quantitatively the effects of a thermal term
of the form $\sim \alpha H^4$ on the late-time era, by analyzing
several observable quantities related directly to the late-time
era. As we showed by using a numerical analysis approach, the
ordinary $f(R)$ gravity and the thermally corrected $f(R)$ gravity
have differences only when the coefficient $\alpha$ of the thermal
corrections takes quite large values, and the differences can be
spotted in statefinders only for redshifts $z\geq 5$, where the
dark energy oscillations are more pronounced in the thermally
corrected $f(R)$ gravity.

Before ending, we need to further clarify some issues. Firstly,
the effect of the thermal terms on the matter and radiation
domination eras. In theory this term should not have any
quantitative role during the matter and radiation domination eras,
since there is no horizon related to these eras. So our strategy
was to add this term by hand and see, just from curiosity, its
effects explicitly. Our results demonstrated that this term, if it
was actually present, it would destroy the phase space structure
of the matter and radiation domination eras, in the context of
$f(R)$ gravity. Thus our analysis showed that indeed, the presence
of the thermal radiation term should not be present during the
matter and radiation domination eras. Initially we thought that it
would not affect at all the dynamical system, but contrary to our
expectation, adopting an unquestionable formal and mathematically
rigorous approach, this term utterly affects the matter and
radiation eras, thus it should not be present during the matter
and radiation domination eras at all.

Another issue is related to the interesting question of having
finite-time singularities related for example in some subspace of
the total phase space, for example in the de Sitter subspace of
trajectories. This question was also addressed by us in Ref.
\cite{Odintsov:2018uaw}, but not for $f(R)$ gravity phase space
and its subspaces. The finite-time singularity analysis of Ref.
\cite{Odintsov:2018uaw} was focused on perfect fluid-related
dynamical systems. But the original question of examining
finite-time singularities in the framework of $f(R)$ gravity phase
space subspaces, is quite deep in its essence. We believe that the
two concepts, namely the finite-time singularities and the fixed
points are not necessarily related, at least at the fundamental
level of the dynamical systems analysis. In a dynamical system,
the phase space and its subspaces are studied, so the presence of
a fixed point means that there is attraction, or repulsion for
this fixed point. A finite time singularity always leads to
blow-ups in the trajectory belonging to the total phase space and
in some specific phase space describing a class of trajectories.

But as we mentioned, the above question is deeper and insightful.
This question was addressed in Ref. \cite{Odintsov:2018uaw} for
perfect fluid dynamical systems, and to our opinion, it might be
possible that a trajectory belonging to a de Sitter, for instance,
subspace of the total phase space, might end up to a sudden
blow-up values for the parameters that quantify the phase space,
if the de Sitter fixed point is unstable for instance. We aim to
formally address this issue in a future work.

\end{document}